\documentclass[conference]{IEEEtran}
\IEEEoverridecommandlockouts
\usepackage{cite}
\usepackage{amsmath,amssymb,amsfonts,amsthm}
\usepackage{algorithm}
\usepackage[noend]{algpseudocode}
\usepackage{graphicx}
\usepackage{textcomp}
\usepackage{xcolor}
\usepackage{xspace}
\usepackage{mathtools}
\usepackage{subcaption}
\usepackage{multirow}
\usepackage{multicol}

\usepackage{tabularx}
\usepackage{arydshln}

\usepackage{url}
\usepackage{array}
\usepackage{tcolorbox}
\usepackage{enumerate}

\usepackage{booktabs}
\usepackage{tabularx}

\newtheorem{theorem}{Theorem}
\newtheorem{definition}{Definition}

\newcommand{\eat}[1]{}

\renewcommand{\color}{\eat}

\newcommand{\ie}{\emph{i.e.,}\xspace}
\newcommand{\aka}{\emph{also known as}\xspace}
\newcommand{\eg}{\emph{e.g.,}\xspace}

    \makeatletter
    \newcommand{\linebreakand}{%
      \end{@IEEEauthorhalign}
      \hfill\mbox{}\par
      \mbox{}\hfill\begin{@IEEEauthorhalign}
    }
    \makeatother

\graphicspath{
    {figures/countermeasures}
    {figures/degree/epsilon}
    {figures/degree/beta}
    {figures/degree/gamma}{figures/clustering_coefficient/epsilon}
    {figures/clustering_coefficient/beta}
    {figures/clustering_coefficient/gamma}
    {figures/illustrations}
    {figures/evaluation_attack/clustering_coefficient}
    {figures/evaluation_attack/modularity}
} 

\def\BibTeX{{\rm B\kern-.05em{\sc i\kern-.025em b}\kern-.08em
    T\kern-.1667em\lower.7ex\hbox{E}\kern-.125emX}}
\begin{document}

\title{Data Poisoning Attacks to Local Differential Privacy Protocols for Graphs
}
\eat{\author{\IEEEauthorblockN{Xi He}
\IEEEauthorblockA{\textit{Faculty of Information Technology} \\
\textit{Macau University of Science and Technology}\\
Macau SAR, China \\
hexichn@gmail.com}
\and
\IEEEauthorblockN{Kai Huang$^{\S}$}
\IEEEauthorblockA{\textit{Faculty of Information Technology} \\
\textit{Macau University of Science and Technology}\\
Macau SAR, China \\
kylehuangk@gmail.com}
\linebreakand
\IEEEauthorblockN{Qingqing Ye}
\IEEEauthorblockA{\textit{Department of Electrical and Electronic Engineering} \\
\textit{The Hong Kong Polytechnic University}\\
Hong Kong SAR, China \\
qqing.ye@polyu.edu.hk}
\and
\IEEEauthorblockN{Haibo Hu}
\IEEEauthorblockA{\textit{Department of Electrical and Electronic Engineering} \\
\textit{The Hong Kong Polytechnic University}\\
Hong Kong SAR, China \\
haibo.hu@polyu.edu.hk}
}}
\author{
 \vspace{0.05in}
 Xi He$^1$, Kai Huang$^{1,\S}$,  Qingqing Ye$^2$,  Haibo Hu$^2$\\
   $^1$ Faculty of Information Technology, Macau University of Science and Technology, Macau SAR \\
    $^2$  The Hong Kong Polytechnic University, Hong Kong SAR \\
 % $^3$  The Hong Kong University of Science and Technology, Hong Kong SAR \\

 \emph{hexichn@gmail.com, kylehuangk@gmail.com, qqing.ye$|$haibo.hu@polyu.edu.hk}
 %\vspace{-0.05in}
}

\maketitle
\begingroup\renewcommand\thefootnote{\textsection}
\footnotetext{Kai Huang is the corresponding author.}
\endgroup

\begin{abstract}
Graph analysis has become increasingly popular with the prevalence of big data and machine learning. Traditional graph data analysis methods often assume the existence of a trusted third party to collect and store the graph data, which does not align with real-world situations. To address this, some research has proposed utilizing Local Differential Privacy (LDP) to collect graph data or graph metrics (e.g., clustering coefficient). This line of research focuses on collecting two atomic graph metrics (the adjacency bit vectors and node degrees) from each node locally under LDP to synthesize an entire graph or generate graph metrics. However, they have not considered the security issues of LDP for graphs.

In this paper, we bridge the gap by demonstrating that an attacker can inject fake users into LDP protocols for graphs and design data poisoning attacks to degrade the quality of graph metrics. In particular, we present three data poisoning attacks to LDP protocols for graphs. As a proof of concept, we focus on data poisoning attacks on two classical graph metrics: degree centrality and clustering coefficient. We further design two countermeasures for these data poisoning attacks. Experimental study on real-world datasets demonstrates that our attacks can largely degrade the quality of collected graph metrics, and the proposed countermeasures cannot effectively offset the effect, which calls for the development of new defenses.
\end{abstract}


\section{Introduction}
Graph analytics has become a powerful tool in various applications such as social networks and finance analysis, enabling the extraction of valuable insights from graph data. However, the sensitive and personal nature of such data raises significant privacy concerns, as exemplified by 
Facebook-Cambridge Analytica data scandal where personal data belonging to millions of Facebook users was collected without their consent to be used for political advertising. This scandal was mainly caused by the Facebook Graph API for third-party apps \cite{stephanie2018,facebook2020}  which allows these apps to access users' friend lists (\ie neighbor lists in a graph) without strict authorizations. Traditional privacy models for graph data, such as k-neighborhood anonymity\cite{zhou2008}, k-degree anonymity\cite{liu2008}, and differential privacy\cite{nissim2007}, rely on a centralized trusted party to manage and release the data. Unfortunately, the centralized party, even Facebook,  cannot be fully trusted to release graph data on behalf of each user. In this context, the problem of \textit{decentralized graphs} has been studied in recent years. In such a scenario,  no party has access to the whole graph, which means that (i) no party can collect the entire graph 
 unfetteredly and derive important graph metrics, and (ii) everyone (\ie a node in the graph) has a local view (\eg number of neighbors and neighbor list) of the graph. This leads to the inability of existing of privacy-preserving graph publication and analysis approaches since the data cannot be collected in the first place.

Local differential privacy (LDP)\cite{duchi2013} has emerged as a promising approach for decentralized graph data privacy. LDP effectively addresses privacy concerns by enabling individual users to perturb their data locally before sharing them, thus eliminating the need for a trusted central party. This approach shifts the control of privacy to the individual user level. LDP has already been successfully applied in various domains, such as frequency estimation and heavy-hitter identification, enabling privacy-preserving data collection and analysis without compromising user privacy. Recent research, including studies  LDPGen\cite{qin2017} and LF-GDPR\cite{ye2020}, has demonstrated the potential of LDP for graph data. These works have shown how LDP can be adapted to preserve the structural properties of graphs while providing strong privacy guarantees. For example, LDPGen proposed to generate synthetic graphs that maintain key topological features of the original graph while satisfying local differential privacy constraints. LF-GDPR introduced a framework for the estimation of privacy-preserving graph metrics under local differential privacy. In both works, users send two atomic metrics, the \textit{adjacency bit vector} and \textit{node degree}, to the
central server (\ie data collector).

Despite the growing adoption of LDP protocols in graph analysis, there is a notable gap in the research concerning the security of these protocols. In addition, there have been studies on the security of LDP protocols for tasks such as frequency estimation and heavy-hitter identification\cite{cao2021}, but \textit{the security of LDP protocols tailored for graph analysis remains unexplored}.

In this paper, we bridge the gap by proposing a family of \textit{data poisoning attacks} to LDP protocols for graph analysis. In our attacks, an attacker can manipulate some users (\aka fake users) to send arbitrary degree values and adjacency bit vectors to the central server, with the goal of preventing the central server from collecting accurate and useful graph metrics of arbitrary attacker-chosen users (\ie attacking targets/nodes in the graph). As a proof of concept, we focus on two graph analysis tasks: \textit{degree centrality} and  \textit{clustering coefficient} estimation. {\color{blue}Degree centrality provides insight into the connectivity and prominence of a node in the network}. 
The clustering coefficient of a node represents the probability that two neighbors of the node are also connected to each other.  The significance of our work lies in the potential consequences of such attacks. For example, consider a social network that employs an LDP protocol to estimate user popularity based on centrality measures. An attacker could manipulate the centrality measures of specific target users, making them appear more popular and influential than they actually are. This manipulation could have far-reaching effects, potentially misleading other users, influencing platform algorithms, and affecting user interactions. %In a more severe scenario, such attacks could be used to artificially inflate the perceived importance of certain accounts, potentially facilitating the spread of misinformation or manipulating public opinion.

Designing effective data poisoning attacks against LDP protocols for graphs introduces non-trivial challenges.
First, as graph data is inherently more complex than values, LDP protocols for graphs often involve complex aggregation techniques, making it more challenging to devise attack strategies that can successfully degrade the accuracy of the collected graph metrics. Second, it is challenging to maximize the attack effectiveness given a limited number of fake users an attacker can inject. To address these challenges, we first introduce the absolute gain (denoted by $Gain$) of an attack that measures the difference of graph metrics after and before the attack. 
By maximizing the absolute gain, we derive the optimized attack called \textit{Maximal Gain Attack} (MGA, see Section \ref{sec:threeattacks}). We also propose two baseline attacks (\ie Random Value Attack (RVA) and Random Node Attack (RNA) in Section \ref{sec:threeattacks}) to better demonstrate the effectiveness of MGA.  In addition, we explore two countermeasures to defend against our attacks to highlight the need for new defenses against our attacks.

Our main contributions are summarized as follows:
\begin{itemize}
    \item We are the first to investigate the security of LDP for graph metrics including degree centrality and clustering coefficient.
    \item We present three data poisoning attacks, including RVA, RNA and MGA, to LDP for graph metrics. We theoretically prove that our  MGA attack can maximally distort the collected graph metrics of targeted nodes. 
    \item We explore two countermeasures against our proposed attacks. 
    \item Experimental
study on real-world datasets demonstrates that our attacks can
largely degrade the quality of collected graph metrics, and the
proposed countermeasures cannot effectively offset the effect,
which calls for the development of new defenses.
\end{itemize}

The remainder of this paper is organized as follows. Section \ref{sec:relatedwork} discusses related work and Section \ref{sec:background} presents the preliminaries. 
In Section \ref{sec:threeattacks}, we present three data poisoning attacks, followed by their detailed implementations on degree centrality (see Section \ref{sec:attackingdegree}) and clustering coefficient (see Section \ref{sec:attackingcluster}).  We present two countermeasures in Section \ref{sec:countermeasures} and experimental results in Section \ref{sec:experiment}. We conclude the paper in Section \ref{sec:conclusion}. Formal proofs of theorems are in \cite{tech}.

\eat{
% 介绍背景
% 1. 介绍图分析的重要性
% 2. 图数据存在隐私保护的需求
% 3. 介绍传统的图数据隐私保护模型及其局限性
Graph analysis has become a cornerstone in various applications, ranging from social network analysis and web graph mining to transportation systems and knowledge bases, driving significant advancements in these fields by extracting meaningful insights from graph data. However, the sensitive nature of graph data, which often includes personal and confidential information, necessitates robust privacy-preserving mechanisms. The Facebook data breach involving Cambridge Analytica serves as a notable example of the risks associated with inadequate privacy measures, where the personal profiles of 87 million Facebook users were harvested through the Facebook Graph API, allowing third-party apps to access users' friends lists with minimal authorization\cite{stephanie2018,facebook2020}. 

Traditional privacy models for graph data, such as k-neighborhood anonymity\cite{zhou2008}, k-degree anonymity\cite{liu2008}, and differential privacy\cite{nissim2007}, typically rely on a centralized trusted party to manage and release the data. While these approaches have shown some efficacy, they are vulnerable to single points of failure and require users to place complete trust in the central authority. This centralized model has become increasingly problematic in an era of growing data breaches and privacy concerns.

% 草稿：另一种写法
% Protecting the privacy of graph data is of paramount importance in today's data-driven world. Local Differential Privacy (LDP) has emerged as a promising solution to safeguard individual privacy without relying on a trusted data curator. In LDP, each user locally perturbs their data before sending it to the data collector, ensuring that privacy is preserved even if the collector is compromised. Two recent LDP protocols, LDPGen and LF-GDPR, have been specifically designed for graph data and aim to address the limitations of previous methods.

% 介绍 Motivation
% 1. LDP 是一种有前景的解决方案
% 2. LDP 已经被用于 Frequency estimation 和 Heavy hitter identification
% 3. 最近也有人将 LDP 用于图数据
Local differential privacy (LDP)\cite{duchi2013} has emerged as a promising alternative for decentralized graph data privacy. LDP effectively addresses privacy concerns by enabling individual users to perturb their data locally before sharing it, thus eliminating the need for a trusted central party. This approach shifts the control of privacy to the individual user level. LDP has already been successfully applied in various domains, such as frequency estimation and heavy hitter identification, enabling privacy-preserving data collection and analysis without compromising user privacy. 

Recent research, including studies like LDPGen\cite{qin2017} and LF-GDPR\cite{ye2020}, has demonstrated the potential of LDP for graph data. These works have shown how LDP can be adapted to preserve the structural properties of graphs while providing strong privacy guarantees. For example, LDPGen proposed a method to generate synthetic graphs that maintain key topological features of the original graph while satisfying local differential privacy constraints. Similarly, LF-GDPR introduced a framework for the estimation of privacy-preserving graph metrics under local differential privacy.

% 草稿：另一种写法
% Prior to LDPGen and LF-GDPR, existing privacy models for graphs assumed a centralized trusted party to release the graph data while satisfying certain privacy metrics, such as k-neighborhood anonymity, k-degree anonymity, k-automorphism, k-isomorphism, and differential privacy. However, in practice, even large companies like Facebook cannot be fully trusted or may not be in a centralized position to release graph data on behalf of each user. Moreover, for decentralized graphs where each user or party maintains a limited view of the graph, there is no central trusted party.

% 介绍 Motivation
% 1. 目前已有针对于用于 Frequency estimation 和 Heavy hitter identification 的 LDP 协议的攻击
% 2. 还没有对与图 LDP 的攻击
% 3. 介绍攻击图 LDP 可能带来的后果
Despite the growing adoption of LDP protocols in graph analysis, there is a notable gap in the research concerning the security of these protocols. Although there have been studies on the security of LDP protocols for tasks such as frequency estimation and heavy hitter identification\cite{cao2021}, the specific security challenges of LDP protocols tailored for graph analysis remain largely unexplored.

% 我们的贡献
% 1. 总的说一下我们做了什么 - 我们对针对于图的 LDP 协议提出了三种攻击方式，在我们的攻击中，攻击者可以构建虚假的用户，并向中心服务器发送任意值的度数和邻接位向量，攻击者的目标是使得中心服务器无法收集准确、有效的数据（图指标）。
% 2. 这件事是有意义的
In this paper, we address this significant gap by proposing three attack strategies targeting LDP protocols for graph analysis. In our attacks, an attacker can create fake users and send arbitrary degree values and adjacency bit vectors to the central server, with the goal of preventing the central server from collecting accurate and useful graph metrics. The significance of our work lies in the potential consequences of such attacks. For example, consider a social network that employs an LDP protocol to estimate user popularity based on centrality measures. An attacker could manipulate the centrality measures of specific target users, making them appear more popular and influential than they actually are. This manipulation could have far-reaching effects, potentially misleading other users, influencing platform algorithms, and affecting user interactions. In a more severe scenario, such attacks could be used to artificially inflate the perceived importance of certain accounts, potentially facilitating the spread of misinformation or manipulating public opinion.

% 这个问题有什么挑战？
% 和向量相比，图数据更加复杂；图指标也更复杂，因为图指标通常用来描述整体特征而非单个节点的特征，因此，用于图的 LDP 协议也更加复杂 -> 想要设计出有效的攻击策略更有挑战。
Designing effective attack strategies against LDP protocols for graph analysis poses unique challenges compared to attacks on LDP for simpler data types like frequency estimation or heavy hitter identification. Graph data is inherently more complex, characterized by intricate relationships and global properties that go beyond individual node characteristics. Consequently, LDP protocols for graphs are more sophisticated, often involving complex aggregation techniques. This complexity makes it more challenging to devise attack strategies that can successfully compromise the accuracy of the collected graph metrics while remaining undetected.

% 简述提出的三种方法
Our proposed attack strategies take into account these unique challenges. The first strategy employs a randomized approach. The second strategy utilizes a more targeted approach. The third strategy, which we term the optimization-based approach, formulates the attack as an optimization problem, seeking to maximize the manipulation of graph metrics.

% 在真实数据集上进行了实验
To evaluate the effectiveness of our proposed attack strategies, we conduct experiments on real-world datasets. Our results show that all three strategies can successfully degrade the accuracy and utility of the graph metrics collected by the central server. Among the three strategies, the optimization-based approach achieves the best performance, while the other randomized and targeted approaches also demonstrate a notable impact on the data accuracy.

Our main contributions are summarized as follows:
\begin{itemize}
    \item We introduce the first data poisoning attack against LDP protocols for graph metric estimation, Specifically targeting two graph metrics: degree centrality and clustering coefficient.
    \item We empirically validate the effectiveness of our attack method. Our experiments demonstrate that the attack can significantly reduce the accuracy of estimated graph metrics.
    \item We explore potential countermeasures against our proposed attack method.
\end{itemize}
}

\section{Related Work}\label{sec:relatedwork}
%The most germane to this work includes local differential privacy, data poisoning attacks,  countermeasures against LDP protocols, and graph privacy.

% 回顾和总结与当前研究相关的已有工作
\textbf{Local Differential Privacy.} 
The concept of Differential Privacy (DP) \cite{dwork2006} was initially introduced for a setting where a trusted data curator collects data from individual users, processes it in a differentially private manner, and releases the results. This approach ensures that the influence of any single element in the dataset on the output is limited. However, in reality, third parties cannot always be trusted. Local differential privacy (LDP) \cite{duchi2013} has been proposed to address this issue. In a local setting, there is no trusted third party. An aggregator aims to collect data from users who are willing to assist the aggregator but do not fully trust the aggregator. To preserve privacy in this scenario, each user perturbs their own data before sending it to the aggregator over a secure channel.

% By randomizing data locally before sharing, LDP limits the aggregator's ability to infer individual users' private data. The aggregator can only access noisy data from each user. Nevertheless, overall statistics about the population can still be estimated through aggregation while providing formal privacy guarantees for each user without relying on a trusted intermediary.

% 目前有关 LDP 的研究
Building on this foundation, a significant body of research has focused on applying LDP, particularly in the areas of frequency estimation and practical data analysis tasks \cite{bassily2017,du2024topk,wang2024ldppurifier,Wang2021Continuous,bassily2015,erlingsson2014,kairouz2014,wang2017,avent2017,cormode2018,jia2019,kairouz2016,qin2016,ren2018,wang2019,wang2018,wang2021,wang2020,ye2022,zhang2018,ye2019,ye2020,ye2021,ren2022,wang2022,zhou2022,zhang2023,du2023,li2023,Mao2024PrivShape,Collaborative2023,Stateful2023Optimized}. This research has led to the development of state-of-the-art protocols for frequency estimation, such as kRR\cite{kairouz2014}, OUE\cite{wang2017}, and OLH\cite{wang2017}. Although these protocols have primarily aimed to enhance data utility within the LDP framework, they have also revealed potential vulnerabilities to data poisoning attacks. Our work is orthogonal to this research. %Our work is orthogonal to this line of research.

\textbf{Data Poisoning Attacks and Countermeasures against LDP Protocols.}
Some researchers have begun to investigate data poisoning attacks against LDP protocols, particularly those used for frequency estimation and heavy-hitter identification. These attacks can be broadly categorized into targeted\cite{cao2021} and untargeted attacks\cite{cheu2021}. Targeted attacks aim to increase the estimated frequencies for items chosen by the attacker (\ie targets). Cao et al.\cite{cao2021} proposed three such strategies: Random perturbed-value attack (RPA), Random item attack (RIA), and Maximal gain attack (MGA). {\color{blue}In contrast, untargeted attacks aim to disrupt the overall frequency distribution rather than targeting specific items. These attacks manipulate the global structure of the data by increasing the distance (often measured by an $L_p$-norm) between the original and manipulated frequency vectors. As a result, these attacks hinder the accurate recovery of the entire data distribution.}

There are several countermeasures against data poisoning attacks on LDP protocols \cite{jagielski2018,mozaffari2015} for machine learning models instead of graph metrics. 
Recently, Cao et al. \cite{cao2021} present three countermeasures against data poisoning attacks to LDP protocols for non-graph data, namely, \textit{normalization}, \textit{fake users detection}, and \textit{conditional probability based detection}. The first one normalizes estimated item frequencies so that each estimated item frequency is nonnegative and the overall item frequency is $1$.  Fake users detection detects potential fake users based on their reported values and removes their impact on the final results. The last one detects fake users based on conditional probability, which is inapplicable when there are more than two target items. The former two countermeasures can take effect in some scenarios but cannot adapt to others and have limited effectiveness in reducing the impacts introduced by data poisoning attacks.
To address this problem, Huang et al. \cite{huang2024} proposes a general countermeasure that can detect fake users in a uniform way and offset their side effects on the quality of collected data. 

However, these attacks and countermeasures are not designed for LDP protocols for graph data.
% Previous LDP research has primarily focused on improving the utility of LDP protocols, leaving a significant gap in studying the security of these protocols. Xiaoyu Cao et al.\cite{cao2021} helped fill this gap by studying data poisoning attacks against LDP protocols specifically used for frequency estimation and heavy hitter identification. They proposed three attack strategies: Random perturbed-value attack (RPA), Random item attack (RIA), and Maximal gain attack (MGA).

\textbf{Privacy-Preserving Graph Release and Processing.} Researchers have developed various approaches to protect sensitive information when releasing graph data \cite{samarati2001,zhou2008,liu2008,zou2009}. However, as these methods have proven vulnerable to de-anonymization \cite{narayanan2009}, more robust privacy concepts have emerged, including differential privacy \cite{nobari2014,dwork2006,sala2011,mir2012,lu2014}. Differential privacy, in particular, has been extensively studied for various graph metrics and statistics \cite{nissim2007,karwa2011,sun2019,kasiviswanathan2013,hay2009,wang2013}. Additionally, significant work has been done on privacy-preserving graph processing methods, including secure shortest path queries \cite{Das2010Anonymizing}, privacy-preserving shortest distance queries \cite{gao2011neighborhood}, subgraph queries \cite{cao2011privacy,Huang2022Privacy}, and general graph queries \cite{Huang2024FRESH}. These studies typically assume the existence of a central server with access to the entire graph. In this paper, we focus on decentralized graphs.

\eat{Researchers have developed various approaches to protect sensitive information when releasing graph data. Early approaches centered on anonymization techniques derived from $k$-anonymity\cite{samarati2001}, such as $k$-neighborhood anonymity\cite{zhou2008}, $k$-degree anonymity\cite{liu2008}, and $k$-automorphism\cite{zou2009}, to counter various structural attacks. However, as these methods proved vulnerable to de-anonymization\cite{narayanan2009}, more robust privacy notions emerged, including $L$-opacity\cite{nobari2014} and differential privacy\cite{dwork2006}. The latter employs generative graph models like \textit{dK-series}\cite{sala2011}, Stochastic Kronecker Graph (SKG)\cite{mir2012}, and Exponential Random Graph Model (ERGM)\cite{lu2014} to fit the original graph and produce synthetic graphs for analysis. Centralized differential privacy, in particular, has been extensively studied for various graph metrics and statistics. Researchers have developed methods to estimate the cost of minimum spanning trees, count triangles\cite{nissim2007}, and perform subgraph counting queries\cite{karwa2011,sun2019} (such as $k$-stars, $k$-triangles, and $k$-cliques) while maintaining privacy. Other work has focused on estimating node degree distributions\cite{kasiviswanathan2013,hay2009} and clustering coefficients\cite{wang2013} under differential privacy constraints.

In addition, there has been a host of work on
privacy-preserving graph processing methods such as secure shortest path queries \cite{Das2010Anonymizing}, privacy-preserving shortest distance queries\cite{gao2011neighborhood},  privacy-preserving subgraph queries \cite{cao2011privacy,Huang2022Privacy}, and general privacy-preserving graph queries \cite{Huang2024FRESH}.  These works assume that there is a central server that has access to the whole graph. In this paper, we focus on decentralized graphs.}

\textbf{Local Differential Privacy for Graph Metric Estimation.} Recent research has shifted towards Local Differential Privacy (LDP) for graph data analysis, enhancing privacy guarantees beyond centralized differential privacy. Sun et al. \cite{sun2019} introduced a method for estimating subgraph counts in decentralized graphs, while LF-GDPR \cite{ye2020} offers a framework for estimating various graph metrics under LDP constraints, building on the earlier LDPGen \cite{qin2017}, which focused on generating synthetic graphs. LF-GDPR allows local collection of atomic metrics, such as adjacency bit vectors and node degrees, to estimate metrics like clustering coefficients and modularity without compromising privacy. However, the security of LDP protocols for graph analysis remains largely unexamined, and our work is the first to investigate the security of LDP in relation to graph metrics, including degree centrality and clustering coefficients.
%While centralized differential privacy has proven effective, recent research has explored the application of Local Differential Privacy (LDP) to graph data analysis, offering even stronger privacy guarantees. Sun et al.\cite{sun2019} propose a method to estimate subgraph counts in a decentralized graph, while LF-GDPR\cite{ye2020} introduces a novel framework to estimate graph metrics while preserving privacy under LDP constraints. This work builds upon the previous study LDPGen\cite{qin2017}, which focused on generating synthetic graphs with LDP guarantees. LF-GDPR addresses the challenge of estimating various graph metrics, such as clustering coefficients and modularity, without compromising individual privacy. The framework provides a parameterized approach that allows for the estimation of different graph metrics by collecting two atomic graph metrics, the adjacency bit vector and node degree, from each node locally. However, the security of LDP protocols tailored for graph analysis remains largely unexplored. We are the first to investigate the security of LDP for graph metrics, including degree centrality and clustering coefficient. 

\section{Preliminaries}\label{sec:background} 
% 为读者提供理解论文主要内容所需的背景知识和基本概念

\newlength{\extraarrayvspace}
\setlength{\extraarrayvspace}{1.2px}

\begingroup
\begin{table}
\caption{List of key notations.}
\centering
\begin{tabularx}{0.48\textwidth}{lX}
\toprule
Notation & Description \\
\midrule
$\varepsilon, \varepsilon_1, \varepsilon_2$ & Privacy budgets\tabularnewline[\extraarrayvspace]
%$n, m, r$ & Number of genuine users, fake users,  target users\tabularnewline[\extraarrayvspace]
$n, m, r$ & \# of genuine users, fake users,  targets\tabularnewline[\extraarrayvspace]
%$r$ & Number of target users\tabularnewline[\extraarrayvspace]
$N = n+m$ & Total number of users\tabularnewline[\extraarrayvspace]
$T = \{t_1, t_2, \ldots, t_r \}$ & Target nodes\tabularnewline[\extraarrayvspace]
$Y = \{y_1, y_2, \ldots, y_m \}$ & Crafted value from fake users\tabularnewline[\extraarrayvspace]
$D=\{d_1, d_2, \ldots, d_N\}$ & Degree vector\tabularnewline[\extraarrayvspace]
$B= \{b_1, b_2, \ldots, b_N \}$ & Adjacency bit vector \tabularnewline[\extraarrayvspace]
$M=\{B_1, B_2, \ldots, B_N \}$ & Adjacency matrix \tabularnewline[\extraarrayvspace]
$\tilde{b}, \tilde{d}$ & Perturbed adjacency bit vector and degree\tabularnewline[\extraarrayvspace]
$Gain$ & Overall gain \tabularnewline[\extraarrayvspace]
%$F$ & Target graph metric \tabularnewline[\extraarrayvspace]
% $\tilde{F}$ & Estimated target graph metric \tabularnewline[\extraarrayvspace]
$\tilde{f}_{i, b}$, $\tilde{f}_{i, a}$  & Estimated metrics of node $i$ before/after attack \tabularnewline[\extraarrayvspace]
$\Delta \tilde{f}_i = \tilde{f}_{i, a} - \tilde{f}_{i, b}$& Change in graph metrics for node $i$ \tabularnewline[\extraarrayvspace]
$c_i$ & Degree centrality for node $i$\tabularnewline[\extraarrayvspace]
$cc_i$ & Clustering coefficient for node $i$\tabularnewline[\extraarrayvspace]
$cc_{B, i}, cc_{A, i}$ & Clustering coefficient for node $i$ before/after attack\tabularnewline[\extraarrayvspace]
$\Delta cc_i = cc_{A, i} - cc_{B, i}$ & Change in clustering coefficient for node $i$\tabularnewline[\extraarrayvspace]
$\mathcal{R}(\cdot)$ & Calibration function \tabularnewline[\extraarrayvspace]
%$\beta$ & The fraction of fake users \tabularnewline[\extraarrayvspace]
%$\gamma$ & The fraction of target users 
$\beta$, $\gamma$ & The fraction of fake users/target users \tabularnewline[\extraarrayvspace]
%$\gamma$ & The fraction of target users \tabularnewline[\extraarrayvspace]
%$V = \{1, 2, \ldots, N\}$ & Set of nodes \tabularnewline[\extraarrayvspace]
%$E \subseteq V \times V$ & Set of edges \tabularnewline[\extraarrayvspace] 
%$G = (V, E)$ & G: Graph, V: Nodes, E: Edges \tabularnewline[\extraarrayvspace]
%$\theta$ & Edge density \tabularnewline[\extraarrayvspace]
$\theta$, $\widetilde{\theta}$ & Edge density of original/perturbed graph \tabularnewline[\extraarrayvspace]
$\tau_i$ & Number of triangles incident to node $i$ \tabularnewline[\extraarrayvspace]
%$\widetilde{\theta}$ & Edge density of the perturbed graph\tabularnewline[\extraarrayvspace]
$p$ & Perturbation probability\tabularnewline[\extraarrayvspace]
\bottomrule
\end{tabularx}
\label{tab:symbols}
\end{table}
\endgroup

Table~\ref{tab:symbols} lists the key notations and acronyms used in this paper. Given a privacy budget $\varepsilon$ ($\varepsilon \geq 0$), the formal definition of $\varepsilon$-LDP is as follows.
\begin{definition}\textup{(\textbf{Local Differential Privacy}).} {\color{blue}
An algorithm $\mathbf{A}$ satisfies $\varepsilon$-local differential privacy ($\varepsilon$-LDP), where $\varepsilon \geq 0$, if and only if for any input $v_1$ and $v_2$, we have
\begin{equation}
\label{eq:2.1}
\forall O \subseteq  \operatorname{Range}(\mathbf{A}): \frac{\operatorname{Pr}\left[\mathbf{A}\left(v_1\right)\in O\right]}{\operatorname{Pr}\left[\mathbf{A}\left(v_2\right) \in O\right]} \leq e^{\varepsilon}
\end{equation}
where $\text{Range}(\mathbf{A})$ denotes the set of all possible outputs of the algorithm $\mathbf{A}$.}
\end{definition}

As a strong privacy measure, LDP has been widely applied in collecting real-value, bit vector, key-value data, and graph data. In this paper, we focus on graph data.
Formally, a graph $G$ is defined as $G = (V, E)$, with $V = \{1, 2, ..., N\}$ representing the set of nodes and $E \subseteq V \times V$ denoting the set of edges. The LDP on the graphs encompasses two primary variants: \textit{Node Local Differential Privacy} (Node LDP) and \textit{Edge Local Differential Privacy} (Edge LDP). Node LDP ensures that the output of a randomized algorithm does not reveal the existence of any individual node in the graph, while edge LDP guarantees that the algorithm's output does not disclose the presence of any specific edge.

\begin{definition}\textup{(\textbf{Node Local Differential Privacy}).} A randomized algorithm $\mathbf{A}$ satisfies $\varepsilon$-node local differential privacy ($\varepsilon$-node LDP), where $\varepsilon \geq 0$, {\color{blue}if and only if for any two adjacency bit vectors (\ie neighbor lists)  $B$ and $B'$}, we have %A randomized algorithm $\mathbf{A}$ satisfies $\varepsilon$-node local differential privacy ($\varepsilon$-node LDP), where $\varepsilon \geq 0$, {\color{blue}if and only if for any two adjacency bit vectors (\ie neighbor lists)\footnote{{\color{blue} An adjacency bit vector is a vector representation of a node's neighbor list. Formally,  a user $u$'s neighbor list can be represented as a $N$-dimensional bit vector $B=(b_1,...,b_N)$, \ie $b_i = 1,i = 1,...,N$, if and only if there is an edge $(u, u_i)$ in the graph; otherwise $b_i = 0$.}}  $B$ and $B'$}, we have
\begin{equation}
\label{eq:nodeldp}
\forall o \in \operatorname{Range}(\mathbf{A}): \frac{\operatorname{Pr}\left[\mathbf{A}\left(B\right)=o\right]}{\operatorname{Pr}\left[\mathbf{A}\left(B'\right)=o\right]} \leq e^{\varepsilon}
\end{equation}
\end{definition}

This definition ensures that the algorithm's output remains approximately the same regardless of the presence or absence of any single node in the graph.

%Edge Local Differential Privacy (Edge LDP) is defined as follows.
\begin{definition}\textup{(\textbf{Edge Local Differential Privacy}).}
A randomized algorithm $\mathbf{A}$ satisfies $\varepsilon$-edge local differential privacy ($\varepsilon$-edge LDP), where $\varepsilon \geq 0$, if and only if for any two adjacency bit vectors $B$ and $B'$ that differ only in one bit, we have
\begin{equation}
\label{eq:edgeldp}
\forall o \in \operatorname{Range}(\mathbf{A}): \frac{\operatorname{Pr}\left[\mathbf{A}\left(B\right)=o\right]}{\operatorname{Pr}\left[\mathbf{A}\left(B'\right)=o\right]} \leq e^{\varepsilon}
\end{equation}
\end{definition}

Observe that edge LDP is a relaxation of node LDP. It limits the definition of neighbors to any two adjacency bit vectors that differ only in one bit (\ie one edge). As edge LDP can still achieve strong indistinguishability of each edge’s existence, in this paper, LDP for graphs refers to edge LDP.

\eat{\color{blue}
\textbf{Remark.} Node-LDP (or Edge-LDP) differs from that of Node-DP (or Edge-DP) as defined in the work \cite{kasiviswanathan2013}.  In graph differential privacy (\ie node-DP and edge-DP), the subject of discussion is a single graph $G$; ``neighboring datasets'' refer to two graphs $G$ and $G'$ that differ in one node (\ie node-DP) or one edge (\ie edge-DP). In contrast, in local differential privacy on graphs, each subject pertains to the private data of a single user, \ie a node in a graph.
}

\subsection{Data Poisoning Attacks on Frequency Estimation}
Data poisoning attacks on machine learning models \cite{Jinyuan2021Intrinsic,Ling2021Adversarial,Minghong2020Influence,Minghong2020Local} are a type of adversarial attack, where the attacker manipulates the training data to cause the model to learn an incorrect or biased function as the attacker desires. This can degrade the model's performance on test data.  On the other hand, data poisoning attacks on frequency estimation (\ie estimating the frequency of an item) aim to increase the estimated frequencies of attacker-chosen items (\ie targets) by injecting fake users into the system. There are three main strategies for data poisoning attacks on frequency estimation under LDP: RPA, RIA, and MGA. RPA selects a value from the encoded space of the LDP protocol uniformly at random for each fake user and sends it to the central server. In contrast to RPA, RIA considers information about the target items. Specifically, RIA randomly chooses a target item from each fake user's set of target items. The LDP protocol then encodes and adds noise to the item. Finally, the perturbed value is sent to the server. The idea behind MGA is to craft the perturbed values for the fake users by solving an optimization problem, with the objective of maximizing the gain of the target item before and after the attack. 

\subsection{Local Differential Privacy for Graph Metric Estimation}

LF-GDPR \cite{ye2020} implements a four-step process for privacy-preserving graph metric estimation. First, it reduces the target graph metric $F$ using a polynomial mapping based on the adjacency matrix $M$ and degree vector $D$. Next, it allocates a privacy budget $\varepsilon$ between the adjacency vector and node degree perturbation to minimize estimation error. In the third step, nodes locally perturb their adjacency and degree values according to the allocated budgets. Finally, the data collector aggregates the perturbed data to estimate the graph metric, applying a calibration function to correct any bias introduced during perturbation. Compared to LF-GDPR, LDPGen \cite{qin2017} is more complicated and ad hoc. In particular,  for different tasks (\eg clustering coefficient estimation), dedicated LDP solutions must be designed from scratch. Therefore, this paper adopts the same approach as LF-GDPR to collect data, including the degree and adjacency bit vector, from each user.

%LF-GDPR \cite{ye2020} implements a four-step process to achieve privacy-preserving graph metric estimation:
\eat{
\begin{itemize}
    \item Graph Metric Reduction: This is the initial step in the framework, designed to express the target graph metric $F$ as a function of the adjacency matrix $M$ and degree vector $D$. This process is formalized as a polynomial mapping function $F = \text{Map}(M, D)$. The key equation for this reduction is:
    \begin{equation}
    F = \sum_l F_l = \sum_l f^{\phi_l}(M^{k_l}) \cdot g^{\psi_l}(D)
    \end{equation}
    where $M^{k_l}$ represents $k_l$-th power of the adjacency matrix, $f^{\phi_l}(\cdot)$ and $g^{\psi_l}(\cdot)$ are aggregation functions applied after projections on the matrix and vector respectively.

    \item Privacy Budget Allocation: The framework optimally divides the total privacy budget $\varepsilon$ between the adjacency bit vector ($\varepsilon_1$) and node degree perturbation ($\varepsilon_2$):
    $\varepsilon_1 = \alpha \varepsilon, \varepsilon_2 = (1-\alpha)\varepsilon$
    where $\alpha$ is determined by minimizing the expected squared error of the estimated metric.

    \item Local Perturbation: Each node locally perturbs its adjacency bit vector $B$ and degree $d$ using the allocated privacy budgets: 
    \begin{equation}
    \widetilde{b}_{i}=\left\{\begin{array}{ll}
    b_{i} & \text { w.p. } \frac{e^{\varepsilon_{1}}}{1+e^{\varepsilon_{1}}} \\
    1-b_{i} & \text { w.p. } \frac{1}{1+e^{\varepsilon_{1}}}
    \end{array} \right.
    \end{equation}
    and 
    \begin{equation}
    \tilde{d} = d + Lap(\frac{2}{\varepsilon_2})
    \end{equation}
    where $Lap(\frac{2}{\varepsilon_2})$ denotes Laplace noise with scale $\frac{2}{\varepsilon_2}$.
    
    \item Aggregation and Calibration: The data collector receives the perturbed adjacency matrix $\widetilde{M}$ and degree vector $\widetilde{D}$, and uses these to estimate the target graph metric $\widetilde{F}$. This estimation is performed using the following equation:
    \begin{equation}
    \widetilde{F} = \sum_l \mathcal{R}(f_{\phi_l}(\widetilde{M}^{k_l})) \cdot g_{\psi_l}(\widetilde{D})
    \end{equation}
    Here, $\mathcal{R}(\cdot)$ is a calibration function designed to suppress the aggregation bias introduced by the perturbation of $\widetilde{M}$. This calibration is necessary for $f_{\phi_l}(\widetilde{M}^{k_l})$ but not for $g_{\psi_l}(\widetilde{D})$, as $\widetilde{D}$ is already an unbiased estimation of $D$ due to the Laplace mechanism used in its perturbation.

    The calibration function $\mathcal{R}$ is derived as a mapping between $f_{\phi_l}(M^{k_l})$ and $f_{\phi_l}(\widetilde{M}^{k_l})$, essentially estimating the true value of $f_{\phi_l}(M^{k_l})$ after observing the perturbed value $f_{\phi_l}(\widetilde{M}^{k_l})$. This relationship is formally expressed as:
    $\mathcal{R}: f_{\phi_l}(\widetilde{M}^{k_l}) \rightarrow f_{\phi_l}(M^{k_l})$. 
\end{itemize}
}
%LF-GDPR and LDPGen \cite{qin2017} are the only two existing works on LDP for graph metric estimation. Compared to LF-GDPR, LDPGen \cite{qin2017} is more complicated and ad hoc. In other words,  for different tasks (\eg clustering coefficient estimation), dedicated LDP solutions must be designed from scratch. Therefore, this paper adopts the same approach as LF-GDPR to collect data, including the degree and adjacency bit vector, from each user.

\section{Data Poisoning Attacks to LDP for Graphs} \label{sec:threeattacks}

\subsection{Threat Model}
It is crucial to understand the potential threats of poisoning attacks to LDP, and the intentions of possible attackers.   In this context, we characterize our threat model by considering the attacker's capabilities, background knowledge, and objectives.

\begin{figure}
    \centering
    \begin{subfigure}[b]{0.96\linewidth}
        \centering
    \includegraphics[width=\linewidth]{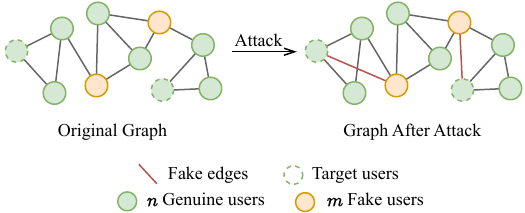}
    \end{subfigure}
    \caption{Data poisoning attack.}
    \label{fig:attacker-in-graph}
\end{figure}

\begin{figure}
    \centering
    \begin{subfigure}[b]{0.98\linewidth}
        \centering
        \includegraphics[width=\linewidth]{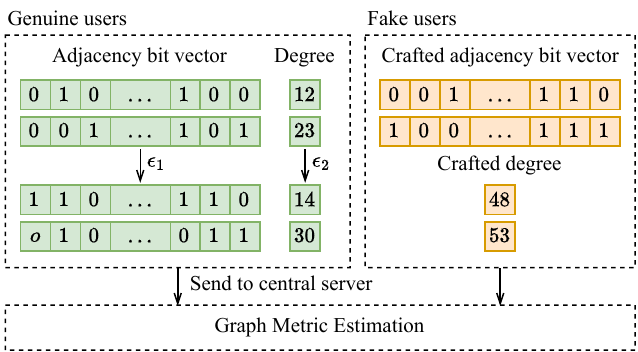}
    \end{subfigure}
    \caption{Genuine users report perturbed adjacency bit vectors and degrees, while fake users report crafted data.}
    \label{fig:genuine-fake-users}
\end{figure}
\textbf{Attacker's capability and background knowledge}. We assume an attacker can control a portion of users in the LDP protocol, and these fake users can send arbitrary data to the central server. As illustrated in Fig. \ref{fig:attacker-in-graph}, {\color{blue}the attacker can manipulate these controlled users to create fake connections within the network.} \footnote{{\color{blue} The paper does not consider the removal of edges in the attacks because removing edges may render the graph disconnected.}} Specifically, we posit that there are $n$ genuine users in the system, while the attacker has control of $m$ fake users. This results in a total of $N=n+m$ users participating in the protocol. The attacker can easily obtain these fake users by compromising existing devices through malware or other means. This is a realistic assumption in many real-world scenarios where user authentication is not stringent or where devices can be infected at scale. Furthermore, attackers can also acquire fake users by purchasing them from underground markets, where large numbers of compromised accounts or devices are often available for sale.

Since the LDP protocol executes the perturbation steps locally on the user's side, the attacker has access to the implementation of these steps. Consequently, the attacker knows various parameters of the LDP protocol. In particular, the attacker is aware of the privacy budget $\varepsilon_1$ used for the adjacency bit vector $B$ and the privacy budget $\varepsilon_2$ used for the degree. Additionally, the attacker knows the domain size for the degree and statistical characteristics of the degree, such as the average degree in the perturbed graph.

\textbf{Attacker's goal}.
We consider the attacker's goal to be degrading the performance of estimating graph metrics for specific target nodes. In other words, the attacker aims to increase the error in the estimated graph metric values for certain target nodes. For instance, some companies may interfere with competitors to cause them to collect inaccurate data. Formally, we denote the set of $r$ target nodes as $T = \{ t_1, t_2, \cdots, t_r \}$. To increase the error in the estimated graph metric values of these target nodes, the attacker carefully crafts the perturbed values sent by fake users to the central server, as illustrated in Fig. \ref{fig:genuine-fake-users}. We can represent the set of crafted values from fake nodes as $Y = \{y_1, y_2, ..., y_m\}$, where $y_i$ denotes the value crafted by the $i$-th fake node. In the context of LDP protocols for graphs, each $y_i$ typically consists of two components: a bit vector representing the node's adjacency information and a value representing the node's degree.

Suppose $\tilde{f}_{t, b}$ and $\tilde{f}_{t, a}$ are the graph metrics estimated by the LDP protocol for a target node $t$ before and after the attack, respectively. We define the gain $\Delta \tilde{f}_t$ for a target node $t$ as
\begin{equation}
\Delta \tilde{f}_t = \lvert \tilde{f}_{t, a} - \tilde{f}_{t, b} \rvert, \forall t \in T
\end{equation}
We then define the overall gain $Gain$ as the sum of gains across all target nodes:
\begin{equation}
Gain = \sum_{t \in T} \Delta \tilde{f}_t
\end{equation}
This overall gain $Gain$ provides a measure of the total impact of the attack on the estimation of graph metrics for all target nodes. A larger value of $Gain$ indicates a more successful attack. Therefore, the attacker's objective is to maximize $Gain$, thereby maximizing the impact of the attack on the target nodes' metric estimations. Formally, the attacker aims to solve the following optimization problem:
\begin{equation}
\label{eq:optimization-problem}
\max_{Y} \quad Gain(Y)
\end{equation}
subject to
\begin{equation}
Y \in \mathcal{Y}
\end{equation}
where $Gain(Y)$ is the overall gain achieved by the set of crafted values $Y$, and $\mathcal{Y}$ is the set of all possible combinations of crafted values constrained by the LDP protocol and the number of fake users $m$.

\subsection{Data Poisoning Attacks}
We propose three attack strategies: Random value attack (RVA), Random node attack (RNA), and Maximal gain attack (MGA). \eat{\footnote{{\color{blue} RVA and RNA are graph-based adaptations of the Random item attack (RIA) and Random perturbed-value attack (RPA) from Cao et al.’s work \cite{cao2021}.}}} These attacks aim to manipulate the estimation of graph metrics for specific target nodes by introducing carefully crafted data from fake nodes. The RVA strategy randomly assigns connections and degree values to fake nodes, RVA does not consider any information about the target nodes. The RNA method focuses on connecting fake nodes to one random target node and then applies normal LDP perturbation to all connections, potentially increasing the impact on specific targets while maintaining plausibility. The MGA crafts the perturbed value for each fake user to maximize the overall gain $Gain$ by solving the optimization problem presented in Equation (\ref{eq:optimization-problem}). RVA and RNA serve as baseline attacks, designed to better demonstrate the effectiveness of MGA.

\begin{itemize}
    \item \textbf{\underline{Random Value Attack (RVA)}}.
This strategy randomly assigns connections and degree values to fake nodes from the entire possible value space, without considering any information about the target nodes.
\item \textbf{\underline{Random Node Attack (RNA)}}. 
This strategy focuses on connecting fake nodes to targets. Specifically, the attacker connects each fake node to one random target node and then applies LDP perturbation to all connections.
\item \textbf{\underline{Maximal Gain Attack (MGA)}}.
This is an optimization-based attack strategy. The attacker crafts the value for each fake user to maximize the overall gain $Gain$ by solving the optimization problem.
\end{itemize}

{\color{blue} Note that RVA (resp. RNA) is a variation of RPA \cite{cao2021} (resp. RIA \cite{cao2021}) adapted for graphs.  In particular, RPA randomly selects a value from the encoded space of the LDP protocol, which, in the context of graphs, translates to randomly assigning connections and degree values to fake nodes from the entire value space. Similarly, RIA randomly selects a target item for each fake user and perturbs that item. When applied to graphs, this corresponds to connecting fake nodes to target nodes and applying LDP perturbation to these connections.
}
%It is worth noting that RVA and RNA are graph-based adaptations of the Random perturbed-value attack (RPA) and Random item attack (RIA) from Cao et al.'s work \cite{cao2021}. The key difference lies in the attack domain: while Cao's work operates on vectors, our graph-based approaches focus on node adjacency vectors and degrees. The similarities remain in their fundamental strategies: RVA and RPA both randomly select from the value space, while RNA and RIA both consider target information and select a single target.

\section{Data Poisoning Attacks against Degree Centrality} \label{sec:attackingdegree}

\begin{figure}
    \centering
    \begin{subfigure}[b]{0.98\linewidth}
        \centering
    \includegraphics[width=\linewidth]{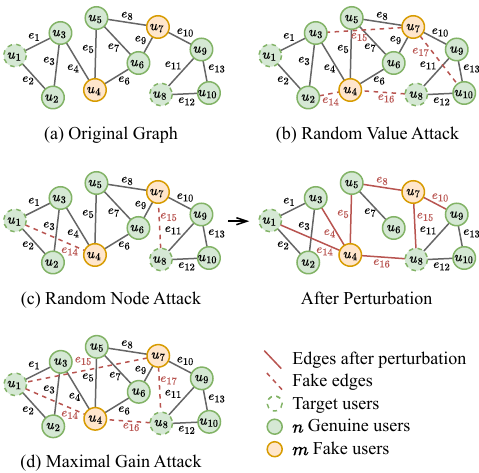}
    \end{subfigure}
    \caption{(a) Original graph; (b)$-$(d) Example of attacking degree centrality using (b) RVA; (c) RNA; and (d) MGA.}
    \label{fig:attack-degree-centrality}
\end{figure}

Degree centrality is a measure of a node's importance in a network based on the number of direct connections it has to other nodes. Formally, the normalized degree centrality of a node $i$ is defined as:
\begin{equation}
c_i = \frac{d_i}{N -1}
\end{equation}
where $d_i$ is the degree of node $i$ and $N$ is the number of nodes. This simple yet powerful metric provides a straightforward way to quantify a node's local influence in the graph. Nodes with high degree centrality are often considered important, as they have many direct connections and potentially greater access to information or resources within the graph.

As shown in Fig. \ref{fig:attack-degree-centrality}, we introduce three attack methods against degree centrality, namely, Random Value Attack (RVA), Random Node Attack (RNA), and Maximal Gain Attack (MGA).

\textbf{Random Value Attack to Degree Centrality}.
To degrade the quality of degree centrality, RVA randomly manipulates the adjacency bit vector of fake nodes in the graph. In particular, it keeps the edges in the original graph unchanged and crafts some edges to be associated with the fake users. In the extreme case, all other nodes are connected with fake users, such that the degree of the fake user is $N-1$. However, the server can easily detect the fake user. To avoid detection due to an excessive number of new connections, the attack limits the number of new connections that can be added. This number is determined by calculating the average degree of all nodes after perturbation based on the adjacency bit vector. 

Fig. \ref{fig:attack-degree-centrality}(b) illustrates an example of the Random Value Attack. Two fake nodes $u_4$ and $u_7$ are randomly connected to other nodes. In particular, node $u_4$ establishes new connections with nodes $u_2$ and $u_8$, while node $u_7$ creates new connections with nodes $u_3$ and $u_{10}$.

\textbf{Random Node Attack to Degree Centrality}. In this attack, each fake node randomly connects to a target node. {\color{blue}Then, the newly created edge, along with the existing connections, is perturbed by an LDP protocol, resulting in a perturbed adjacency bit vector for the fake node.}

%The adjacency list preserves existing connections in the graph. Each fake node is randomly connected to one target node and then normal perturbation is applied to all connections.

As illustrated in Fig. \ref{fig:attack-degree-centrality}(c), there are two fake nodes, $u_4$ and $u_7$. First, each fake node establishes a connection with a single target node. Specifically, $u_4$ forms a new connection with the target node $u_1$, while $u_7$ connects to the target node $u_8$. Then, the crafted connections together with existing edges are perturbed by the LDP protocol. As a result of this perturbation, node $u_4$ gets an additional connection $e_{16}$. In addition, two connections, $e_6$ and $e_9$, are removed from the graph.

\textbf{Maximal Gain Attack to Degree Centrality}. To maximize the gain, each fake node randomly connects to as many target nodes as possible. Then, the newly created edge, along with the existing connections, can be encoded into the adjacency bit vector which is directly sent to the data collector. However, this may expose the identity of a fake node, since all fake nodes may connect with the same set of nodes. Therefore, the number of additional connections for each fake node is carefully controlled. The upper limit for new connections is determined by the average degree of the graph after perturbation, which is calculated based on the $\varepsilon$ value.

Formally, the objective of MGA can be expressed as maximizing the overall gain in degree centrality for all target nodes:
\begin{equation}
\max \sum_{t \in T} \Delta c_t
\end{equation}
where $\Delta c_t$ is the change in degree centrality for target node $t$ after the attack. Given that degree centrality is simply the degree of a node, we can represent $\Delta c_t$ as:
\begin{equation}
\Delta c_t = \frac{1}{N-1}\sum_{u \in U} x_{ut}
\end{equation}
In this equation, $x_{ut}$ is a binary variable that takes the value $1$ if fake node $u$ is connected to target node $t$, and $0$ otherwise.

The most effective way to increase overall gain is to maximize the number of connections between fake nodes and target nodes, subject to the constraints imposed by the average degree limit. Each new connection from a fake node to a target node directly increases the degree of the target node by one, thus contributing to the maximization of the sum of $\Delta c_t$ across all target nodes.

% The upper limit of the degree that this attack can bring to a target node can be expressed as:
% \begin{equation}
% \text{Upper Limit} = d_t + \min(m, \bar{d}')
% \end{equation}
% where $m$ is the number of fake nodes and $\bar{d}'$ is the average degree of the graph after perturbation.

Fig. \ref{fig:attack-degree-centrality}(d) is an example of Maximal Gain Attack. To maximize connections between the target nodes and fake nodes, the fake node $u_4$ simultaneously creates connections with two target nodes, $u_1$ and $u_8$. Similarly, the fake node $u_7$ also forms connections with these two target nodes.

\begin{theorem} 
	{The overall gain of MGA to degree centrality is 
 \begin{equation}
 \text{Gain} = \frac{m \cdot r}{N-1} \cdot \bigg(\frac{\min(r, \lfloor\overline{\tilde{d}}\rfloor)}{r} - \frac{\overline{\tilde{d}}}{N-1}\bigg)
 \end{equation}
 where $\overline{\tilde{d}}$ is the average degree of the graph after perturbation.}
\end{theorem}
\eat{
\begin{proof}
We prove this theorem through the following steps:
\begin{enumerate}
    \item The probability of an existing connection between any two nodes is $p = \frac{\tilde{d}}{N-1}$.
    \item The number of existing connections between fake nodes and target nodes is approximately:
$E_{\text{existing}} \approx m \cdot r \cdot p = m \cdot r \cdot \frac{\tilde{d}}{N-1}$.
    \item Each fake node can add at most $\min(r, \lfloor\tilde{d}\rfloor)$ connections, but we need to subtract the existing connections. So, the maximum number of new connections per fake node is $\min(r, \lfloor\tilde{d}\rfloor) - p \cdot r$.
    \item The total number of new connections that can be added by all fake nodes is:
$E_{\text{new}} = m \cdot (\min(r, \lfloor\tilde{d}\rfloor) - p \cdot r)$
$= m \cdot (\min(r, \lfloor\tilde{d}\rfloor) - \frac{r\tilde{d}}{N-1})$.
    \item The change in degree centrality for a single node when its degree increases by $1$ is $\frac{1}{N-1}$.
    \item Therefore, the overall gain in degree centrality for all target nodes is $\text{Gain} = \frac{E_{\text{new}}}{N-1}$ $= \frac{m}{N-1} \cdot (\min(r, \lfloor\tilde{d}\rfloor) - \frac{r\tilde{d}}{N-1})$ $= \frac{m \cdot r}{N-1} \cdot (\frac{\min(r, \lfloor\tilde{d}\rfloor)}{r} - \frac{\tilde{d}}{N-1})$.
\end{enumerate}
% Let $k=1$, the original top-k edge-diversified patterns discovery problem (TED Problem) becomes to find a single edge-diversified pattern, \ie a graph $g$ that covers maximum number of edges. The reformulated problem (denoted by Simplified TED Problem),  \ie TED Problem with $k=1$,  can be reduced from the maximum coverage problem \cite{Feige1998}, which is a classical NP-hard optimization problem.
\end{proof}
}

\section{Data Poisoning Attacks to Clustering Coefficient}\label{sec:attackingcluster}

The clustering coefficient of a node in a graph measures the degree to which the node's neighbors are connected to each other. Formally, the local clustering coefficient for a node $i$ is defined as:
\begin{equation}
cc_i = \frac{2\tau_i}{d_i(d_i-1)}
\end{equation}
where $\tau_i$ is the number of triangles connected to node $i$, and $d_i$ is the degree of node $i$. A high clustering coefficient indicates that the node's neighbors are densely interconnected.

\textbf{Random Value Attack to Clustering Coefficient}.
In this attack, the adjacency bit vector and degree of fake nodes are manipulated randomly. These fake nodes are connected to arbitrary other nodes in the graph. The newly created connections, combined with the existing connections, form the adjacency bit vector for the fake node. To avoid detection, the number of connections for each fake node is determined based on the average degree of real nodes after perturbation for each $\varepsilon$ value. These connections are not perturbed after being established. The degree value for each fake node is randomly selected from the entire degree space.

\textbf{Random Node Attack to Clustering Coefficient}.
In this attack, each fake node randomly connects to a target node. {\color{blue}Then, the newly created edge, along with the existing connections, is perturbed by an LDP protocol, resulting in a perturbed adjacency bit vector for the fake node.} The degree of each fake node is calculated based on its connections and then perturbed according to the LDP protocol.

\begin{figure}
    \centering
    \begin{subfigure}[b]{0.98\linewidth}
        \centering
        \includegraphics[width=\linewidth]{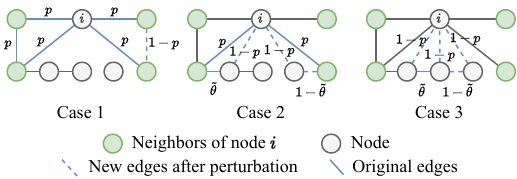}
    \end{subfigure}
    \caption{Number of triangles incident to node $i$.}
    \label{fig:estimate-triangle}
\end{figure}

\textbf{Maximal Gain Attack against Clustering Coefficient}. 
For each fake node, MGA crafts its perturbed value by solving an optimization problem. The change in clustering coefficient for node $i$ before and after the attack can be expressed as:
\begin{equation} \Delta cc_i = cc_{A,i} - cc_{B,i} \end{equation}
where $cc_{A,i}$ and $cc_{B,i}$ represent the clustering coefficients after and before the attack, respectively.
The optimization objective is to maximize the change for each target node:
\begin{equation}
\max \sum_{t \in T} \Delta cc_t
\end{equation}
For the perturbed graph, we can estimate the clustering coefficient of the $i$-th node using the following formula:
\begin{equation}
\displaystyle{c c_i=\frac{2 \tau_i}{\widetilde{d_i}\left(\widetilde{d}_i-1\right)}}
\end{equation}
where $\tau_i$ is the corrected number of triangles and $d_i$ is the perturbed degree. To estimate $\tau_i$, we consider three cases based on the relationships of the other two nodes in the triangle, following the approach proposed in the LF-GDPR\cite{ye2020}, as shown in Fig. \ref{fig:estimate-triangle} where $p$ is the perturbation probability.

\begin{figure}
    \centering
    \begin{subfigure}[b]{0.75\linewidth}
        \centering
    \includegraphics[width=\linewidth]{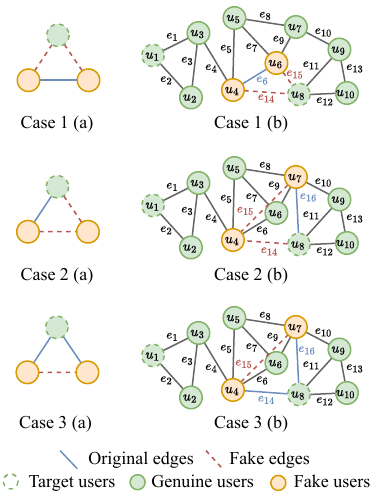}
    \end{subfigure}
    \caption{Connections between fake nodes and target nodes.}
    \label{fig:fake-node-connection}
\end{figure}

\begin{itemize}
    \item Both Nodes are Neighbors (Case 1, Fig. \ref{fig:estimate-triangle}). If there is an edge between the two neighboring nodes, the probability of keeping the triangle in the perturbed graph is $p^3$. Otherwise, the probability of forming the triangle in the perturbed graph is $p^2(1-p)$. Therefore, the total number of such triangles in the perturbed graph is $\tilde{\tau}_{i,1} = \tau_i \cdot p^3 + \left(\frac{1}{2}d(d-1) - \tau_i\right) \cdot p^2(1-p)$.
    \item Only One Node is a Neighbor (Case 2, Fig. \ref{fig:estimate-triangle}). The number of possible triangles is $d(N-d-1)$ and the probability of retaining the triangle in the perturbed graph is $p(1-p)\tilde{\theta}$. Thus, the total number of such triangles in the perturbed graph is $\tilde{\tau}_{i,2} = d(N-d-1) \cdot p(1-p)\tilde{\theta}$.
    \item Neither Node is a Neighbor (Case 3, Fig. \ref{fig:estimate-triangle}). The number of possible triangles is $\frac{1}{2}(N-d-1)(N-d-2)$ and the probability of forming the triangle in the perturbed graph is $(1-p)^2\tilde{\theta}$. Therefore, the total number of such triangles in the perturbed graph equals $\tilde{\tau}_{i,3} = \frac{1}{2}(N-d-1)(N-d-2) \cdot (1-p)^2\tilde{\theta}$.
\end{itemize}
By summing up the results from the three cases, we derive the corrected number of triangles incident to node $i$:
\begin{equation}
\begin{aligned}
\tau_{i} =\mathcal{R}(\widetilde{\tau}_{i})= & \frac{1}{p^2(2 p-1)}(\widetilde{\tau}_{i} -\frac{1}{2} \widetilde{d_i}(\widetilde{d_i}-1) p^2(1-p) \\
& -\widetilde{d_i}(N-\widetilde{d_i}-1) p(1-p) \widetilde{\theta} \\
& -\frac{1}{2}(N-\widetilde{d_i}-1)(N-\widetilde{d_i}-2)(1-p)^2 \widetilde{\theta})
\end{aligned}
\end{equation}
Here, $\widetilde{\theta}$ represents the edge density of the perturbed graph, which can be calculated as:
\begin{equation}
\displaystyle{\widetilde{\theta}=\frac{\sum_{i=1}^N \widetilde{\tau}_i}{N(N-1)}}
\end{equation}
The number of triangles for a certain node before the attack can be expressed as:
\begin{equation}
\begin{aligned}
\tau_{B, i} =\mathcal{R}(\widetilde{\tau}_{B, i})= & \frac{1}{p^2(2 p-1)}(\widetilde{\tau}_{B, i} -\frac{1}{2} \widetilde{d_i}(\widetilde{d_i}-1) p^2(1-p) \\
& -\widetilde{d_i}(N-\widetilde{d_i}-1) p(1-p) \widetilde{\theta} \\
& -\frac{1}{2}(N-\widetilde{d_i}-1)(N-\widetilde{d_i}-2)(1-p)^2 \widetilde{\theta})
\end{aligned}
\end{equation}
Since only the number of triangles changes for a node before and after the attack, while the degree remains constant, the number of triangles after the attack is 
\begin{equation}
\begin{aligned}
\tau_{A, i} =\mathcal{R}(\widetilde{\tau}_{A, i})= & \frac{1}{p^2(2 p-1)}(\widetilde{\tau}_{A, i} -\frac{1}{2} \widetilde{d_i}(\widetilde{d_i}-1) p^2(1-p) \\
& -\widetilde{d_i}(N-\widetilde{d_i}-1) p(1-p) \widetilde{\theta} \\
& -\frac{1}{2}(N-\widetilde{d_i}-1)(N-\widetilde{d_i}-2)(1-p)^2 \widetilde{\theta})
\end{aligned}
\end{equation}
Substitute $\tau_{A,i}$ and $\tau_{B,i}$ into $\Delta cc_i$
\begin{equation}
\Delta cc_i = \frac{2\tau_{A,i}}{\widetilde{d_i}(\widetilde{d_i}-1)} - \frac{2\tau_{B,i}}{\widetilde{d_i}(\widetilde{d_i}-1)}
\end{equation}
\begin{equation}
= \frac{2(\tau_{A,i} - \tau_{B,i})}{\widetilde{d_i}(\widetilde{d_i}-1)}
\end{equation}
\begin{equation}
= \frac{2}{p^2(2p - 1)} \cdot \frac{1}{\widetilde{d_i}(\widetilde{d_i}-1)} \cdot (\tilde{\tau}_{A,i} - \tilde{\tau}_{B,i}) 
\end{equation}
The first two terms in the formula, $\frac{2}{p^2(2p - 1)}$ and $\frac{1}{\widetilde{d_i}(\widetilde{d_i}-1)}$, remain constant before and after the attack. Therefore, the objective is to maximize $\tilde{\tau}_{A,i}$. To achieve this, we discuss several possible connection cases between the target nodes and fake nodes, as shown in Fig. \ref{fig:fake-node-connection}.

\begin{itemize}
    \item \textbf{Neither fake node is connected to the target node (see Case 1a, Fig. \ref{fig:fake-node-connection})}. To form a triangle, each fake node should craft an edge to connect with the target. For example, as shown in Fig. \ref{fig:fake-node-connection} (Case 1b), fake nodes $u_4$ and $u_6$ are already connected to each other via edge $e_6$, but neither is connected to the target node. In this case, we can connect both fake nodes to the target node by creating new edges $e_{14}$ and $e_{15}$ to form a triangle.
    \item \textbf{Only one fake node is connected to the target node (Case 2a, Fig. \ref{fig:fake-node-connection})}. 
    In this case, the attack needs to create two new edges to craft a triangle. In particular, the attack should craft an edge between the target and another fake node that is disconnected from the target. In addition, a crafted edge should be inserted to connect two fake nodes. For example, as shown in Fig. \ref{fig:fake-node-connection} (Case 2b), the target node $u_8$ is connected to a fake node. To create a triangle, it creates the edge $e_{14}$ between the fake node $u_4$ and target $u_8$, and edge $e_{15}$ between the two fake nodes.
    \item \textbf{Both fake nodes are connected to the target node (Case 3a, Fig. \ref{fig:fake-node-connection})}. In such a case, the attack needs to create one new edge between the two fake nodes to craft a triangle. As shown in Fig. \ref{fig:fake-node-connection} (Case 3b), the target node $u_8$ is connected to both fake nodes via edges $e_{14}$ and $e_{16}$, but the fake nodes are not connected to each other. Therefore, the triangle can be constructed by creating the edge $e_{15}$ between the two fake nodes.
\end{itemize}

Observe that the attack can introduce more triangles associated with the target to the graph by crafting edges as above. However, if all fake nodes connect to target nodes,  they may show some patterns that will expose the identity of a fake node. In other words, the fake node can be easily detected. To maximize the gain while disguising the identities of fake nodes, we limit the number of new connections each fake node can create to the average degree calculated from the perturbed graph for each $\varepsilon$ value. To make better use of the limited number of connections, we design a prioritized allocation mechanism where fake nodes are first connected to each other, and only then to target nodes. This strategy helps to introduce more triangles. Once created, these connections remain unperturbed. The degree of each fake node is then calculated based on its connections and subsequently perturbed according to the LDP protocol.

\begin{theorem} 
	{Overall gain of MGA to clustering coefficient is 
 \begin{align*}
\text{Gain} &= r \cdot \frac{2}{p^2(2p - 1)} \cdot \frac{1}{\overline{\widetilde{d}}(\overline{\widetilde{d}}-1)} \\
&\cdot \frac{m}{2 \cdot p'(1-p')^2 + p'^2(1-p')+3\cdot(1-p')^3}
\end{align*}
where $\overline{\tilde{d}}$ is the average perturbed degree and $p' = \frac{\overline{\tilde{d}}}{N-1}$ is the probability of forming a connection.}
\end{theorem}

% The upper limit of the number of triangles that this attack can bring to a target node can be expressed as:
% \begin{equation}
% \text{Upper Limit} = \tau_{B,i} + \min(\binom{m}{2}, \binom{\widetilde{d_i}}{2} - \tau_{B,i})
% \end{equation}
% where $\tau_{B,i}$ is the original number of triangles for the target node before the attack and $\widetilde{d_i}$ is the perturbed degree of the target node.

\section{Countermeasures}\label{sec:countermeasures}
To further validate the effectiveness of the proposed data poisoning attacks, we explore two countermeasures to address these attacks. The first countermeasure aims to detect fake nodes by analyzing the frequent itemsets. The intuition is that attackers may utilize similar patterns to create multiple fake nodes, which could manifest themselves as frequent itemsets in the perturbed bit vectors. The second countermeasure compares the difference between the degree calculated from the perturbed bit vector and the directly reported degree. The idea is that the discrepancy between these two values could serve as an indicator of the presence of fake nodes. As will be seen from experiments on real-world datasets (see Section \ref{sec:experiment}), we find that these countermeasures are ultimately not effective in mitigating the impact of the attacks. The attackers are able to evade detection by the proposed methods, which highlights the need for new defenses against our data poisoning attacks.

% We explored two countermeasures. The first countermeasure detects fake nodes through frequent itemsets, as attackers may use similar patterns to create multiple fake nodes, which might manifest as frequent itemsets in bit vectors. The second compares the difference between the degree calculated from the perturbed bit vector and the directly sent degree. Through experiments on real-world datasets, we found that these countermeasures were not effective in mitigating the impact of attacks.

\subsection{Frequent Itemsets based Detection}
This countermeasure is adapted from the detection method proposed by Cao et al.\cite{cao2021}. It is particularly effective against MGA (Maximal gain attack), as MGA tends to create connections between fake nodes and target nodes, as well as among the fake nodes themselves. This behavior pattern manifests as frequent itemsets in the bit vectors. Therefore, an intuitive countermeasure is to analyze the frequency of frequent itemsets in the bit vectors. The detection method works as follows:

\begin{enumerate}
    \item Utilize the Apriori algorithm\cite{agrawal1994}, a frequent itemset mining technique, to identify connections that appear frequently in the bit vectors, i.e., frequent itemsets.
    \item If the number of frequent itemsets in a node's bit vector exceeds a predefined threshold among the collected bit vectors, that node is classified as a fake node.
    \item Unlike Cao's method\cite{cao2021}, which directly removes detected fake nodes, our approach attempts to reconstruct the previous connections. We reconstruct the previous connections of these fake nodes based on the bit vector information of the real nodes connected to the identified fake nodes.
\end{enumerate}

However, this method has a notable drawback: it may produce false positives, as certain legitimate social patterns can also form frequent itemsets.

\subsection{Degree based Detection}
This countermeasure is particularly effective against Random Value Attack. The principle behind this method is based on the difference between the reported degree and the estimated degree calculated from the perturbed bit vector.

In RVA, the attacker sends a randomly chosen degree from the degree space, which often differs significantly from the estimated degree that can be calculated from the perturbed bit vector. In contrast, genuine nodes typically have a reported degree that follows a Laplace distribution centered around the estimated degree from their perturbed bit vector. The detection method works as follows:  1) for each node, calculate the degree based on the uploaded perturbed bit vector; 2) compare this estimated degree with the directly reported degree. If the difference between these two values exceeds a predefined threshold, mark the node as a fake node; and 3) for each detected fake node, remove its connections from the nodes it claims to be connected to, thereby restoring the degrees of genuine nodes.

\eat{
\begin{enumerate}
    \item For each node, calculate the degree based on the uploaded perturbed bit vector.
    \item Compare this estimated degree with the directly reported degree. If the difference between these two values exceeds a predefined threshold, mark the node as a fake node.
    \item For each detected fake node, remove its connections from the nodes it claims to be connected to, thereby restoring the degrees of genuine nodes.
\end{enumerate}
}

The threshold for detection is set to the maximum degree calculated from the reported perturbed bit vectors plus three times the standard deviation $3\sigma$ of the Laplace distribution.

\section{Evaluation}\label{sec:experiment}
{
\begin{table}
\caption{Datasets.}
\centering
\begin{tabularx}{0.45\textwidth}{llX}
\toprule
Dataset & Number of vertices & Number of edges \\
\midrule
$Facebook$ &  4,039 & 88,234\\
$Enron$ & 36,692 & 183,831\\
$AstroPh$ & 18,772 & 198,110 \\
$Gplus$ & 107,614 & 12,238,285 \\
\bottomrule
\end{tabularx}
\label{tab:dataset_stat}
\end{table}
}

\begin{table}
\caption{Default parameter settings.}
\centering
\begin{tabularx}{0.45\textwidth}{llX}
\toprule
Parameter & Default setting & Description \\
\midrule
$\beta$ & 0.05 & The fraction of fake users\\
$\gamma$ & 0.05 & The fraction of target users\\
$\varepsilon$ & 4 & Privacy budget \\
\bottomrule
\end{tabularx}
\label{tab:parameters-1}
\end{table}

\subsection{Experimental Setup}
In this section, we evaluate the effectiveness of our proposed attacks. All experiments were conducted using Java 17.0.11.

% Performance Measures
\textbf{Datasets}. We evaluate our attacks on four real-world datasets that were widely used in the literature \cite{ye2020}. The statistics are in Table \ref{tab:dataset_stat}.

\begin{itemize}
    \item Facebook: an undirected social network consisting of 4,039 nodes and 88,234 edges, obtained from a survey of participants using the Facebook app.
   %  \item Enron: an undirected email communication network of 36,692 nodes and 183,831 edges. This dataset is derived from the email correspondence of Enron Corporation employees.
    \item Enron: an undirected email communication network of 36,692 nodes and 183,831 edges. This dataset is derived from the email correspondence of Enron Corporation.
    \item AstroPh: an undirected collaboration network of 18,772 authors and 198,110 edges indicating collaborations between authors in arXiv, who submitted papers to the Astro Physical category.
    \item Gplus: an undirected social network of 107,614 users and 12,238,285 edges indicating shares of social circles.
   % \item Gplus: an undirected social network of 107,614 Google+ users and 12,238,285 edges indicating shares of social circles.
\end{itemize}

\textbf{Parameter settings}. For graph metric estimation, the overall gains of our attacks may depend on $\beta$ (\ie the fraction of fake users), $\gamma$ (\ie the fraction of target users), and $\varepsilon$ (\ie privacy budget). Table \ref{tab:parameters-1} shows the default settings, which will be used in our experiments unless otherwise specified. We will also analyze the impact of each parameter while ﬁxing the remaining parameters to their default settings.

\subsection{Results for Degree Centrality Estimation}

% Figure 1 illustrate the impact of different parameters on the overall gains of three attack methods for degree centrality estimation. Across all scenarios, MGA consistently demonstrates superior performance compared to RVA and RNA, with RVA generally outperforming RNA except in some cases. These trends remain consistent across the different social graphs, although the magnitude of gains varies slightly.

\begin{figure}
    \centering
    % First row
    \begin{subfigure}[b]{0.49\linewidth}
        \centering
        \includegraphics[width=\linewidth]{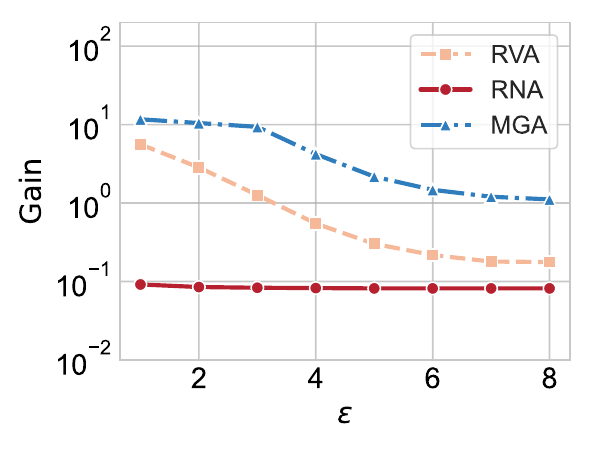}
        \caption{Facebook}
    \end{subfigure}
    \hfill
    \begin{subfigure}[b]{0.49\linewidth}
        \centering
        \includegraphics[width=\linewidth]{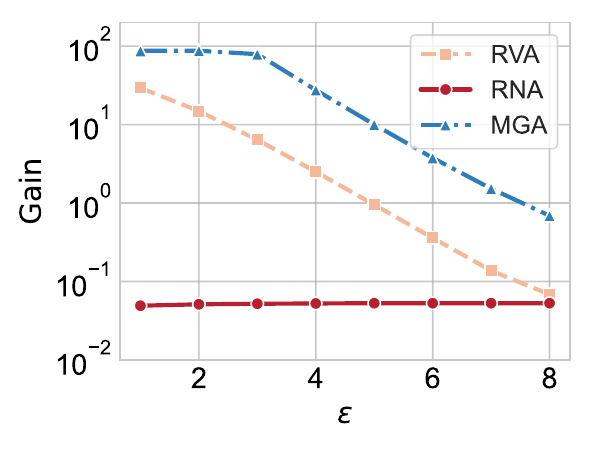}
        \caption{Enron}
    \end{subfigure}

    \vspace{0.5em} 
    % Second row
    \begin{subfigure}[b]{0.49\linewidth}
        \centering
        \includegraphics[width=\linewidth]{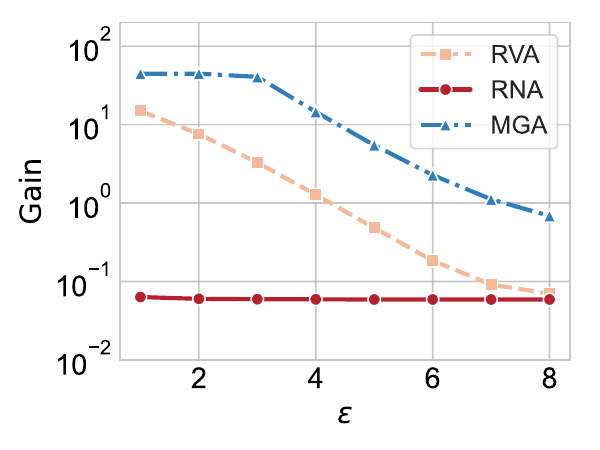}
        \caption{AstroPh}
    \end{subfigure}
    \hfill
    \begin{subfigure}[b]{0.49\linewidth}
        \centering
        \includegraphics[width=\linewidth]{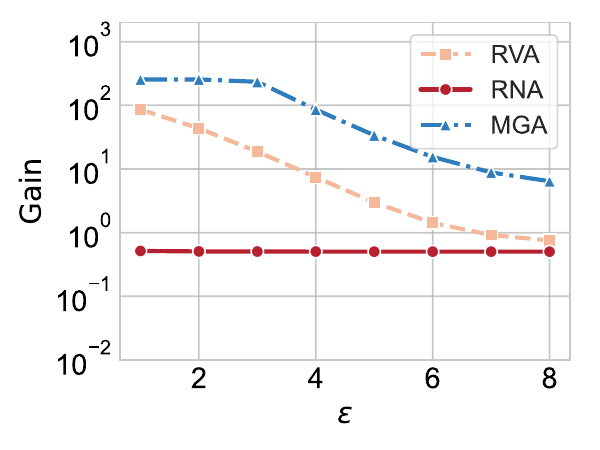}
        \caption{Gplus}
    \end{subfigure}
    
    \caption{Overall gains of attacks to degree centrality.}
    \label{fig:degreeattack_epislon}
\end{figure}

\noindent\underline{\textbf{Exp 1. Overall Results on Degree Centrality}}. We first evaluate the effectiveness of different attack strategies (RVA, RNA, and MGA) on degree centrality estimation under varying privacy budgets. The results are reported in Fig. \ref{fig:degreeattack_epislon} where the privacy budget $\varepsilon$ ranges from $1$ to $8$. As can be seen in the figure, MGA and RVA show a clear inverse relationship between the privacy budget and attack effectiveness. This is because a larger privacy budget introduces a smaller number of edges allowed to be injected, which further leads to less gain. In contrast, the gain of MGA remains nearly unchanged. The reason lies in that RNA connects to only one target node, making it insensitive to changes in the privacy budget. However, no matter what the privacy budget is, MGA always outperforms RVA and RNA, as MGA can connect the limited number of injected edges to the targets. 

\begin{figure}
    \centering
    % First row
    \begin{subfigure}[b]{0.49\linewidth}
        \centering
        \includegraphics[width=\linewidth]{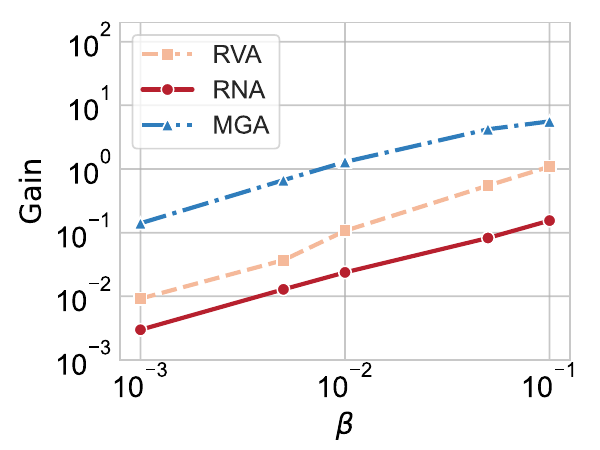}
        \caption{Facebook}
    \end{subfigure}
    \hfill
    \begin{subfigure}[b]{0.49\linewidth}
        \centering
        \includegraphics[width=\linewidth]{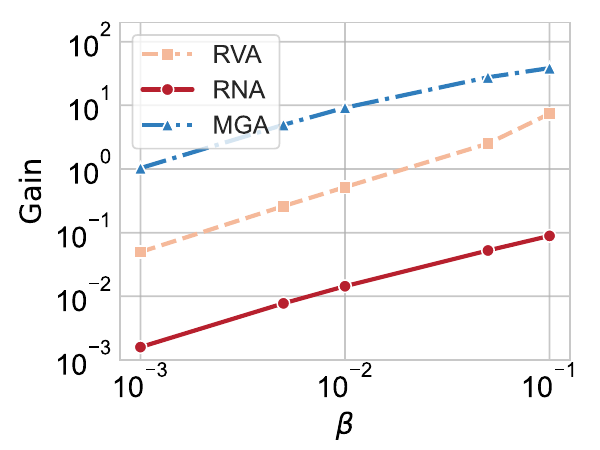}
        \caption{Enron}
    \end{subfigure}

    \vspace{0.5em} 
    % Second row
    \begin{subfigure}[b]{0.49\linewidth}
        \centering
        \includegraphics[width=\linewidth]{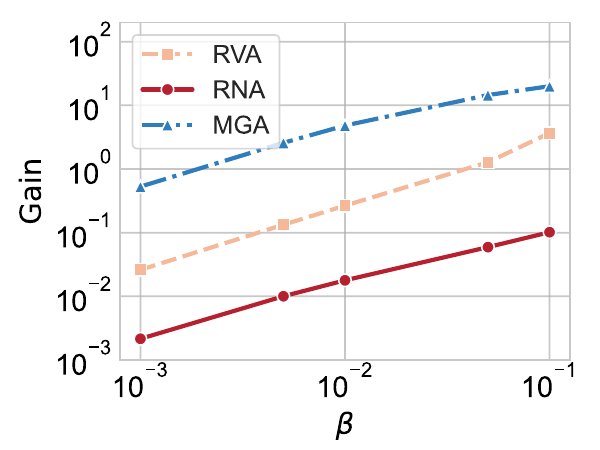}
        \caption{AstroPh}
    \end{subfigure}
    \hfill
    \begin{subfigure}[b]{0.49\linewidth}
        \centering
        \includegraphics[width=\linewidth]{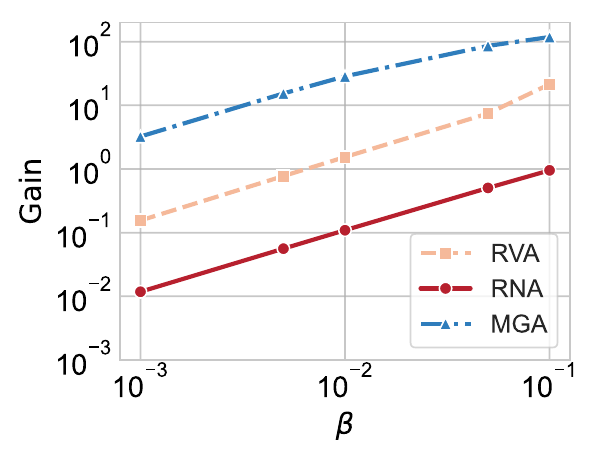}
        \caption{Gplus}
    \end{subfigure}
    
  %  \caption{Impact of $\beta$ on the effectiveness of attacks to degree centrality.}
  \caption{Impact of $\beta$ on attacks to degree centrality.}
    \label{fig:degreeattack_beta}
\end{figure}

\noindent\underline{\textbf{Exp 2. Effect of $\beta$ on Degree Centrality}}. We evaluate the impact of proportions of fake users on the effectiveness of different attack strategies (RVA, RNA, and MGA) for degree centrality estimation. The results are presented in Fig. \ref{fig:degreeattack_beta} where $\beta$ takes values of $0.001$, $0.005$, $0.01$, $0.05$, and $0.1$. As shown in the figure, all three attack methods demonstrate a positive correlation between $\beta$ and attack effectiveness. This trend can be attributed to the increased opportunities for manipulation that a higher number of fake nodes provides, which becomes more pronounced as $\beta$ increases. With more fake users, attackers can create more connections and perturb the graph structure more significantly, leading to a greater impact on degree centrality estimates. Notably, across all values of $\beta$, we observe a consistent performance ranking: MGA consistently outperforms RVA, which in turn outperforms RNA. The reason is that MGA is an optimization-based method, while RNA simply adds one new connection per fake user.

\begin{figure}
    \centering
    % First row
    \begin{subfigure}[b]{0.49\linewidth}
        \centering
        \includegraphics[width=\linewidth]{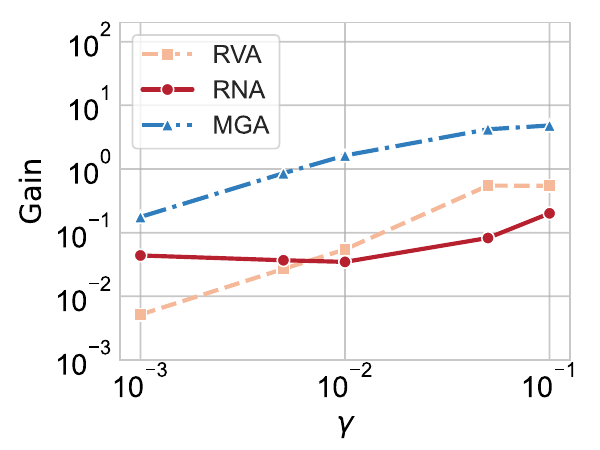}
        \caption{Facebook}
    \end{subfigure}
    \hfill
    \begin{subfigure}[b]{0.49\linewidth}
        \centering
        \includegraphics[width=\linewidth]{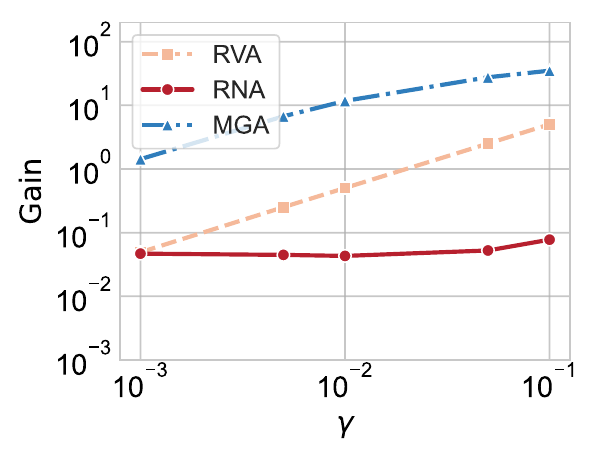}
        \caption{Enron}
    \end{subfigure}

    \vspace{0.5em} 
    % Second row
    \begin{subfigure}[b]{0.49\linewidth}
        \centering
        \includegraphics[width=\linewidth]{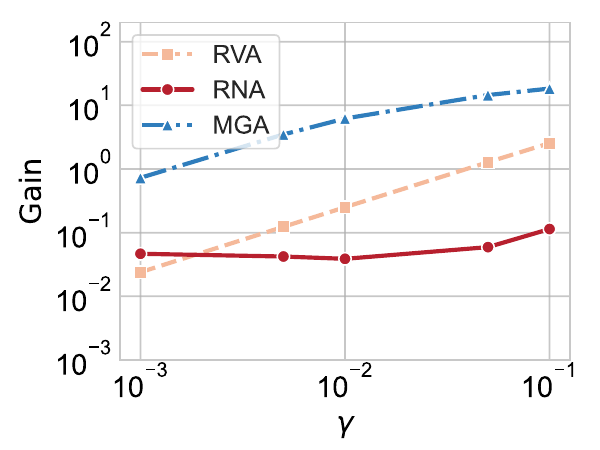}
        \caption{AstroPh}
    \end{subfigure}
    \hfill
    \begin{subfigure}[b]{0.49\linewidth}
        \centering
        \includegraphics[width=\linewidth]{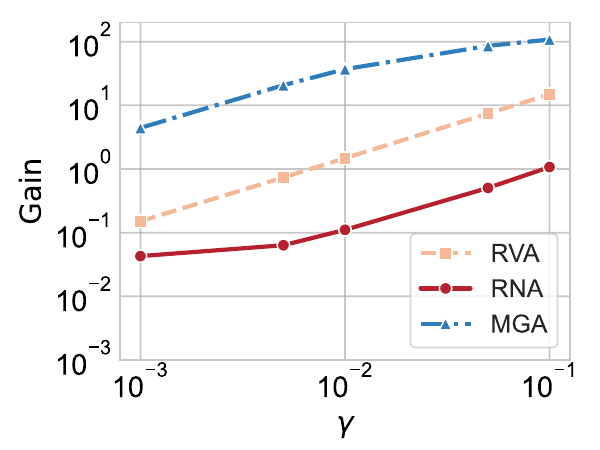}
        \caption{Gplus}
    \end{subfigure}
    
   % \caption{Impact of $\gamma$ on the effectiveness of attacks to degree centrality.}
   \caption{Impact of $\gamma$ on attacks to degree centrality.}
    \label{fig:degreeattack_gamma}
\end{figure}

\noindent\underline{\textbf{Exp 3. Effect of $\gamma$ on Degree Centrality}}. We evaluate the impact of varying the proportion of target nodes on the effectiveness of different attack strategies (RVA, RNA, and MGA) for degree centrality estimation. The results are presented in Fig. \ref{fig:degreeattack_gamma} where $\gamma$ takes values of $0.001$, $0.005$, $0.01$, $0.05$ and $0.1$. As shown in the figure, all three attack methods demonstrate a positive correlation between $\gamma$ and attack effectiveness. This phenomenon can be attributed to the expanded attack surface that a larger number of target nodes provides. With more target nodes available, attackers have a greater pool of potential connections to utilize, allowing them to more effectively alter the graph structure and influence degree centrality estimates. While RNA also shows a generally positive trend, its correlation is less pronounced. Similarly, across all values of $\gamma$, MGA consistently outperforms RVA and RNA. And in the majority of cases, RVA exhibits better performance than RNA.

\begin{figure}
    \centering
    % First row
    \begin{subfigure}[b]{0.49\linewidth}
        \centering
        \includegraphics[width=\linewidth]{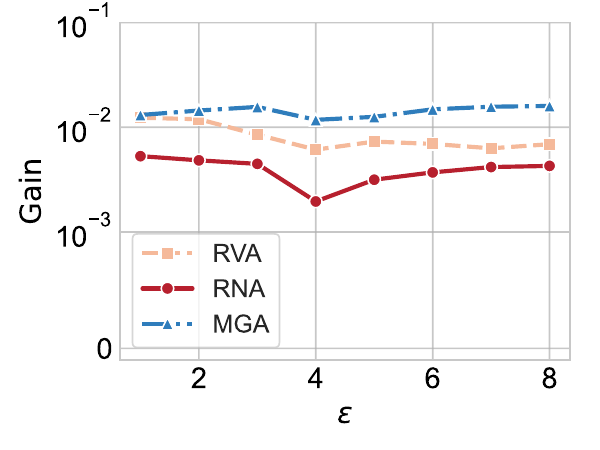}
        \caption{Facebook}
    \end{subfigure}
    \hfill
    \begin{subfigure}[b]{0.49\linewidth}
        \centering
        \includegraphics[width=\linewidth]{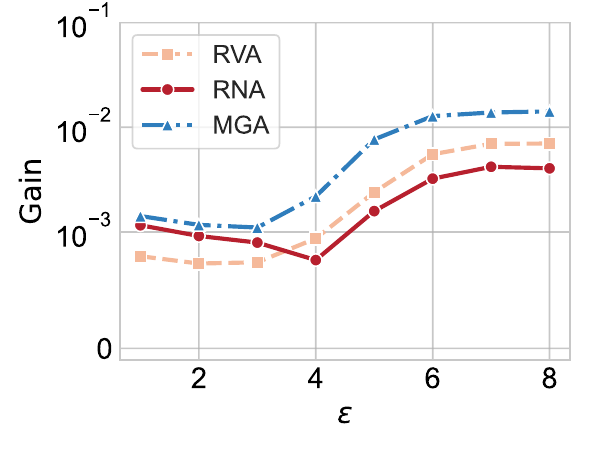}
        \caption{Enron}
    \end{subfigure}

    \vspace{0.5em} 
    % Second row
    \begin{subfigure}[b]{0.49\linewidth}
        \centering
        \includegraphics[width=\linewidth]{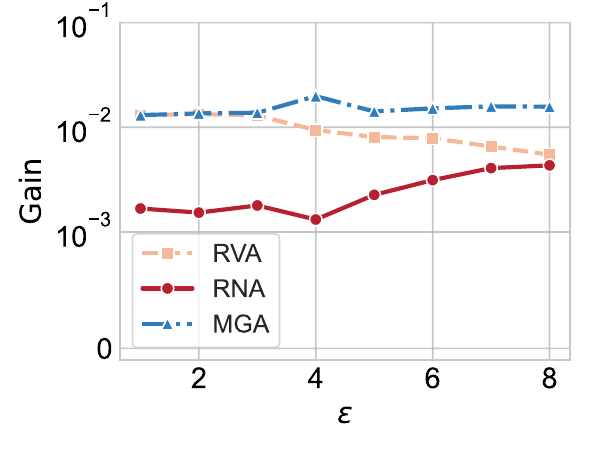}
        \caption{AstroPh}
    \end{subfigure}
    \hfill
    \begin{subfigure}[b]{0.49\linewidth}
        \centering
        \includegraphics[width=\linewidth]{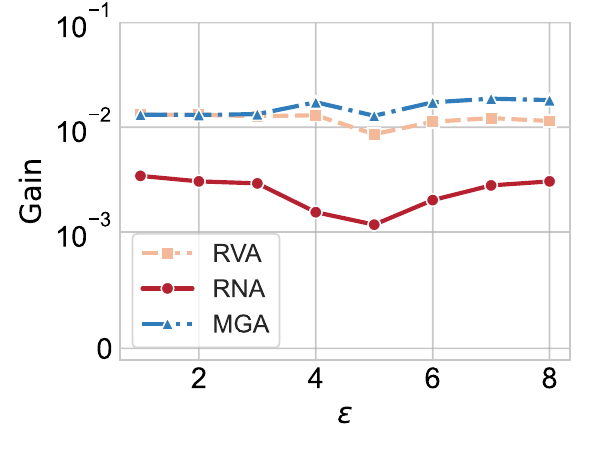}
        \caption{Gplus}
    \end{subfigure}
    
    \caption{Overall gains of attacks to clustering coefficient.}
    \label{fig:clustering_coefficient_attack_epsilon}
\end{figure}

\subsection{Results for Clustering Coefficient Estimation}

\noindent\underline{\textbf{Exp 4. Overall Results on Clustering Coefficient}}. We then evaluate the effectiveness of different attack strategies (RVA, RNA, and MGA) on clustering coefficient estimation under varying privacy budgets. The results are reported in Fig. \ref{fig:clustering_coefficient_attack_epsilon} where the privacy budget $\varepsilon$ also ranges from $1$ to $8$.  The results show that the MGA consistently outperforms both RVA and RNA across all values of $\varepsilon$, with its performance remaining relatively stable as $\varepsilon$ increases. This superior performance can be attributed to its optimization-based approach, which strategically crafts fake user data to maximize the overall gain. Generally, RVA performs better than RNA. The effectiveness of attacks on clustering coefficient estimation appears relatively insensitive to changes in the privacy budget $\varepsilon$ for most datasets, with only slight fluctuations observed. This suggests that for clustering coefficient estimation, the privacy budget does not significantly alter the vulnerability of the graph to these attacks. %However, the specific characteristics of social graphs can influence the impact of the privacy budget on attack effectiveness.

\begin{figure}
    \centering
    % First row
    \begin{subfigure}[b]{0.49\linewidth}
        \centering
        \includegraphics[width=\linewidth]{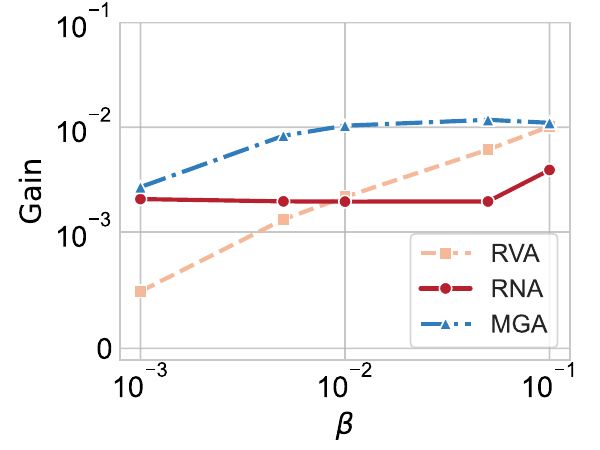}
        \caption{Facebook}
    \end{subfigure}
    \hfill
    \begin{subfigure}[b]{0.49\linewidth}
        \centering
        \includegraphics[width=\linewidth]{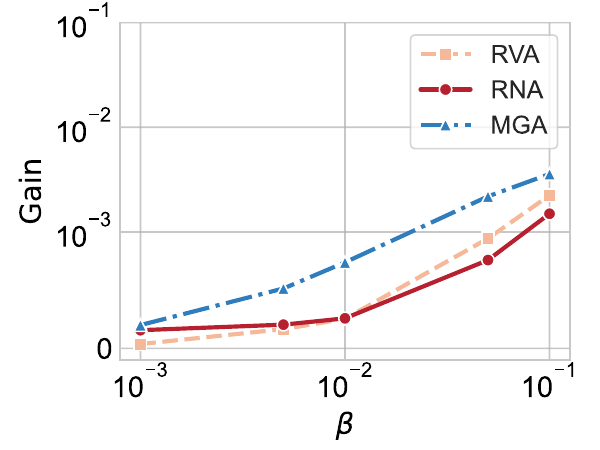}
        \caption{Enron}
    \end{subfigure}

    \vspace{0.5em} 
    % Second row
    \begin{subfigure}[b]{0.49\linewidth}
        \centering
        \includegraphics[width=\linewidth]{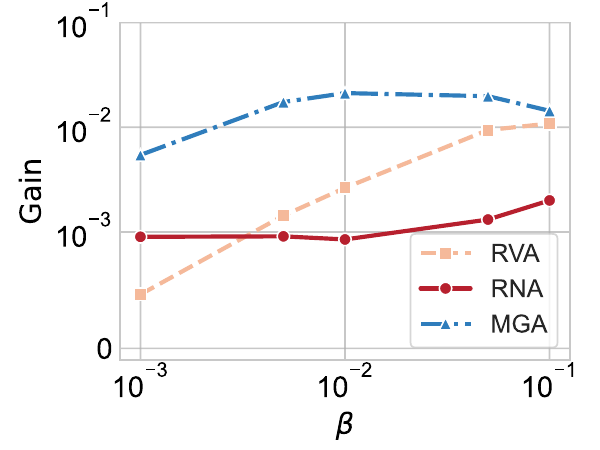}
        \caption{AstroPh}
    \end{subfigure}
    \hfill
    \begin{subfigure}[b]{0.49\linewidth}
        \centering
        \includegraphics[width=\linewidth]{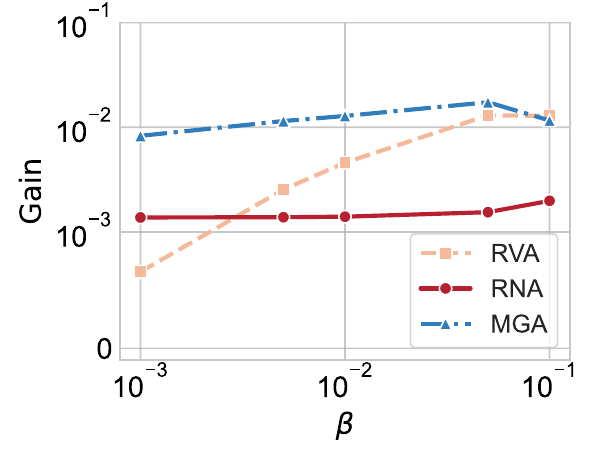}
        \caption{Gplus}
    \end{subfigure}
        
    \caption{Impact of $\beta$ on the effectiveness of attacks to clustering coefficient.}
    \label{fig:clustering_coefficient_attack_beta}
\end{figure}

\noindent\underline{\textbf{Exp 5. Effect of $\beta$ on Clustering Coefficient}}. We evaluate the impact of varying proportions of fake users on the effectiveness of different attack strategies (RVA, RNA, and MGA) for clustering coefficient estimation. The results are reported in Fig. \ref{fig:clustering_coefficient_attack_beta} where the proportion of fake users $\beta$ takes values of $0.001$, $0.005$, $0.01$, $0.05$ and $0.1$. As can be seen in the figure, these attacks demonstrate a positive correlation between $\beta$ and attack effectiveness. This phenomenon can be attributed to the increased number of fake nodes available to the attacker. With more fake users, attackers can create a larger number of fake connections, allowing for more significant manipulation of the graph structure and, consequently, the estimated clustering coefficients. Observe that when $\beta$ reaches a certain threshold (around 0.05-0.1 depending on the dataset), MGA's performance begins to plateau and becomes similar to RVA. This occurs because, at this point, the fake nodes in MGA have already connected to all target nodes. This plateau suggests an upper limit to the effectiveness of MGA when the proportion of fake users becomes sufficiently large.

\begin{figure}
    \centering
    % First row
    \begin{subfigure}[b]{0.49\linewidth}
        \centering
        \includegraphics[width=\linewidth]{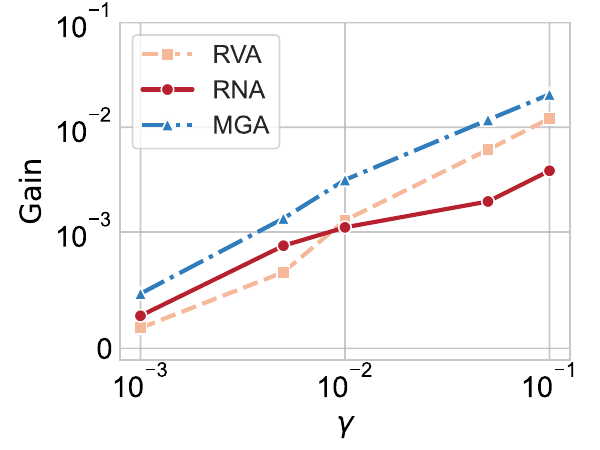}
        \caption{Facebook}
    \end{subfigure}
    \hfill
    \begin{subfigure}[b]{0.49\linewidth}
        \centering
        \includegraphics[width=\linewidth]{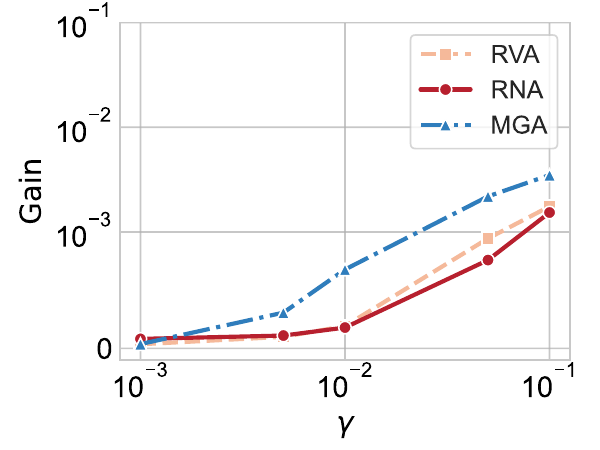}
        \caption{Enron}
    \end{subfigure}

    \vspace{0.5em} 
    % Second row
    \begin{subfigure}[b]{0.49\linewidth}
        \centering
        \includegraphics[width=\linewidth]{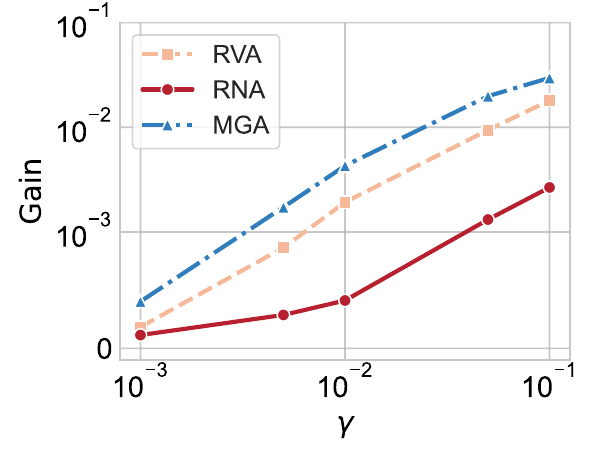}
        \caption{AstroPh}
    \end{subfigure}
    \hfill
    \begin{subfigure}[b]{0.49\linewidth}
        \centering
        \includegraphics[width=\linewidth]{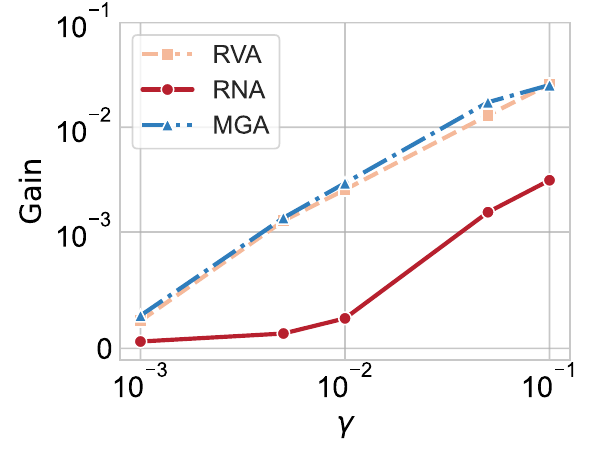}
        \caption{Gplus}
    \end{subfigure}
        
    \caption{Impact of $\gamma$ on the effectiveness of attacks to clustering coefficient.}
    \label{fig:clustering_coefficient_attack_gamma}
\end{figure}

\noindent\underline{\textbf{Exp 6. Effect of $\gamma$ on Clustering Coefficient}}. 
We evaluate the impact of varying the proportion of target nodes on the effectiveness of different attack strategies (RVA, RNA, and MGA) for clustering coefficient estimation. The results are reported in Fig. \ref{fig:clustering_coefficient_attack_gamma} where the proportion of target nodes $\gamma$ takes values of $0.001$, $0.005$, $0.01$, $0.05$, and $0.1$. As can be seen in the figure, all three attack methods demonstrate a positive correlation between $\gamma$ and attack effectiveness. This trend can be attributed to the fact that more target nodes likely imply a greater number of interconnections among these nodes, providing more opportunities for manipulation. MGA consistently outperforms RVA and RNA across all $\gamma$ values and datasets, and RVA shows the second-best performance. We also observe that the relative performance of the attacks and their sensitivity to $\gamma$ varies somewhat across different social graph datasets. This suggests that the underlying structure of the social graph influences the effectiveness of these attacks.

\begin{figure}
    \centering
    % First row
    \begin{subfigure}[b]{0.49\linewidth}
        \centering
        \includegraphics[width=\linewidth]{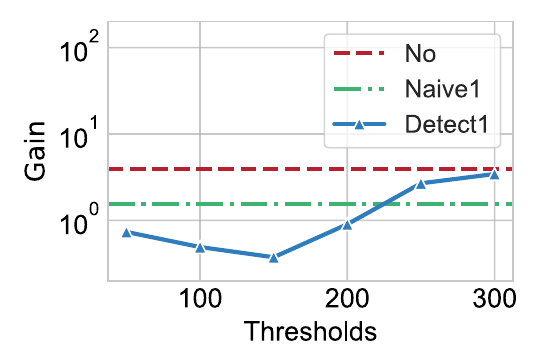}
    \caption{{\color{blue}Countermeasure against MGA}}
    \end{subfigure}
    \hfill
    \begin{subfigure}[b]{0.49\linewidth}
        \centering
        \includegraphics[width=\linewidth]{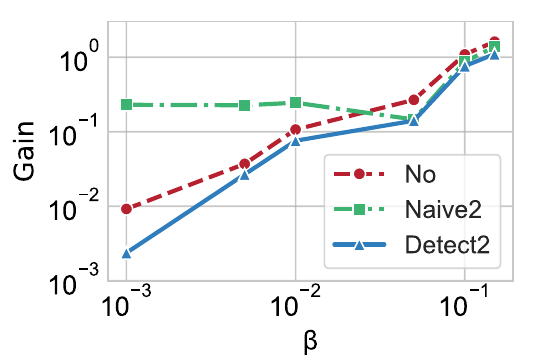}
    \caption{{\color{blue}Countermeasure against RVA}}
    \end{subfigure}
    \caption{{\color{blue}Countermeasures against attacks to degree centrality.}}
\label{fig:countermeasures_degree}
\end{figure}

\subsection{Results for Countermeasures}
\noindent\underline{\textbf{Exp 7. Countermeasures for Degree Centrality}}. We evaluate the effectiveness of two countermeasures against different attack strategies on degree centrality estimation. The results are reported in Fig. \ref{fig:countermeasures_degree}. {\color{blue}To better evaluate the effectiveness of frequent itemsets based detection (denoted by \textsf{Detect1}) in defending against MGA on degree centrality, we implement a naive method (denoted by \textsf{Naive1}) that identifies the top 3\% of highest-degree nodes as attackers.} In Fig. \ref{fig:countermeasures_degree}(a), we vary the detection threshold for the frequent itemsets based method against MGA, with thresholds of $50$, $100$, $150$, $200$, $250$, and $300$. As can be seen in the figure, there is a U-shaped relationship between the detection threshold and the attack gain. At low thresholds, many genuine nodes are mistakenly identified as fake, leading to a decrease in gain. As the threshold increases, fewer nodes are flagged as fake, consequently increasing the attack's impact. This suggests an optimal threshold to maximize the effectiveness of this countermeasure. {\color{blue} Note that \textsf{Detect1} is generally more effective than \textsf{Naive1}, especially when the threshold is not significantly larger than the ratio of fake users.
}

{\color{blue} To better evaluate the effectiveness of our degree based detection method (denoted by \textsf{Detect2}) in defending against RVA on degree centrality, we implement a naive method (denoted by \textsf{Naive2}) that identifies the 
nodes with degrees in either the top or bottom 3\% of the degree distribution as fake nodes.} In Fig. \ref{fig:countermeasures_degree}(b), we evaluate degree-based detection against RVA under different proportions of fake nodes, with $\beta$ values of $0.001$, $0.005$, $0.01$, $0.05$, $0.1$, and $0.15$. As shown in the figure, varying $\beta$ has minimal effect on the performance of the detection method. This is because the random nature of RVA leads to a high false-negative rate when identifying fake nodes, resulting in many fake nodes being missed by the detection algorithm regardless of their proportion. {\color{blue} Compared to \textsf{Detect2}, \textsf{Naive2} exhibits worse efficiency, generating even higher gains than the original results without detection. This is because it tends to classify many genuine nodes as fake.}

%{\color{blue}To better evaluate the effectiveness of countermeasure against RVA (denoted by detect) on degree based detection for degree centrality, we implement a naive method where nodes whose degrees fall within either the top or bottom 3\% of the degree distribution are classified as fake nodes (denoted by naive).} In Fig. \ref{fig:countermeasures_degree}(b), we evaluate degree-based detection against RVA under different proportions of fake nodes, with $\beta$ values of $0.001$, $0.005$, $0.01$, $0.05$, $0.1$, and $0.15$. As shown in the figure, varying $\beta$ has minimal effect on the performance of the detection method. This is because the random nature of RVA leads to a high false-negative rate when identifying fake nodes, resulting in many fake nodes being missed by the detection algorithm regardless of their proportion. {\color{blue}Notably, when $\beta$ is low, the naive method shows even higher gain than the scenario without detection, indicating that it mistakenly identifies many genuine nodes as fake ones.}

%These results demonstrate that although both countermeasures show some effectiveness, neither completely neutralize the attacks.

\begin{figure}
    \centering
    % First row
    \begin{subfigure}[b]{0.49\linewidth}
        \centering
        \includegraphics[width=\linewidth]{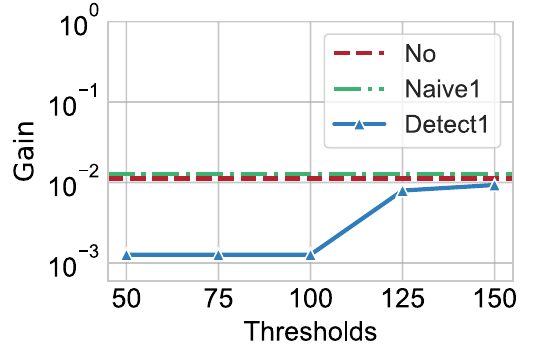}
        \caption{{\color{blue}Countermeasure against MGA}}
        \label{fig:countermeasures-cc-frequent-itemsets}
    \end{subfigure}
    \hfill
    \begin{subfigure}[b]{0.49\linewidth}
        \centering
        \includegraphics[width=\linewidth]{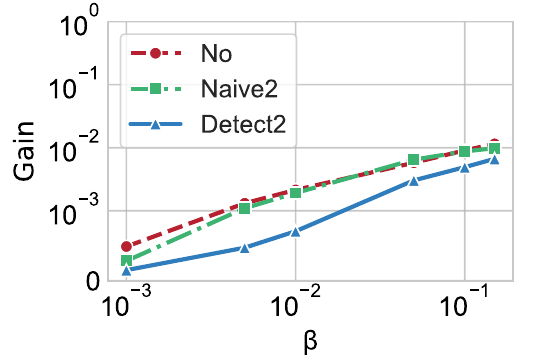}
        \caption{{\color{blue}Countermeasure against RVA}}
        \label{fig:countermeasures-cc-degree}
    \end{subfigure}
    \caption{{\color{blue}Countermeasures against attacks to clustering coefficient.}}
\label{fig:countermeasures_clustering_coefficient_estimation}
\end{figure}

\noindent\underline{\textbf{Exp 8. Countermeasures for Clustering Coefficient}}. We also evaluate the effectiveness of two detection methods against different attack strategies on clustering coefficient estimation. In Fig. \ref{fig:countermeasures_clustering_coefficient_estimation}(a), we vary the detection threshold for the frequent itemsets based method against MGA, using threshold values of $50$, $75$, $100$, $125$, and $150$. As can be seen in the figure, the overall gain initially remains constant as the detection threshold increases, before gradually rising. This trend can be attributed to the threshold's impact on the number of detected fake nodes. At lower thresholds, the method effectively identifies and mitigates the impact of fake nodes. However, as the threshold increases, fewer nodes are flagged as fake, consequently allowing the attack's impact to grow.  In Fig. \ref{fig:countermeasures_clustering_coefficient_estimation}(b), we evaluate degree-based detection against RVA under different proportions of fake nodes, with $\beta$ values of $0.001$, $0.005$, $0.01$, $0.05$, $0.1$, and $0.15$. As shown in the figure, the overall gain after applying the detection method is lower compared to the attack without detection, indicating that the countermeasure does provide some protection against RVA. However, the effectiveness of the detection remains relatively consistent across different $\beta$ values, with only minimal variations observed as $\beta$ changes. This suggests that while the degree-based detection method can reduce the impact of RVA, its performance is not significantly influenced by the proportion of fake nodes in the network. Similarly, while both countermeasures show some effectiveness in mitigating attacks, neither completely neutralizes them. This highlights the need for new defenses against the proposed attacks. {\color{blue} Note that our countermeasures demonstrate greater effectiveness than naive solutions (e.g., \textsf{Detect1} vs. \textsf{Naive1} and \textsf{Detect2} vs. \textsf{Naive2}).  }

\begin{figure}
    \centering
    % First row
    \begin{subfigure}[b]{0.49\linewidth}
        \centering
        \includegraphics[width=\linewidth]{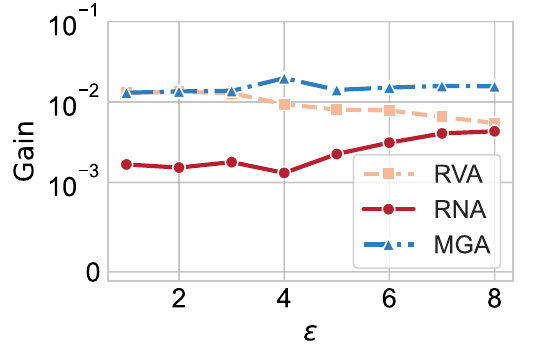}
        \caption{{\color{blue}LF-GDPR}}
    \end{subfigure}
    \hfill
    \begin{subfigure}[b]{0.49\linewidth}
        \centering
        \includegraphics[width=\linewidth]{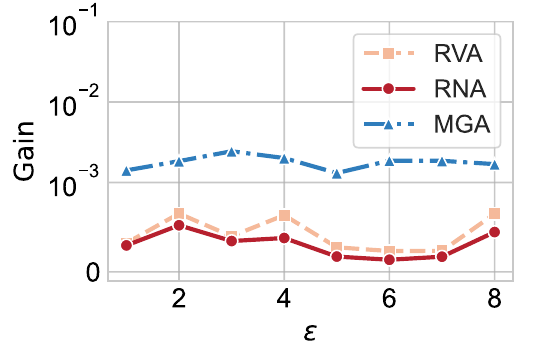}
        \caption{{\color{blue}LDPGen}}
    \end{subfigure}
    
    \caption{{\color{blue}Evaluation of attacks on LF-GDPR and LDPGen for clustering coefficient.}}
    \label{fig:evaluation_attacks_clustering_coefficient}
\end{figure}

\begin{figure}
    \centering
    \begin{subfigure}[b]{0.49\linewidth}
        \centering
        \includegraphics[width=\linewidth]{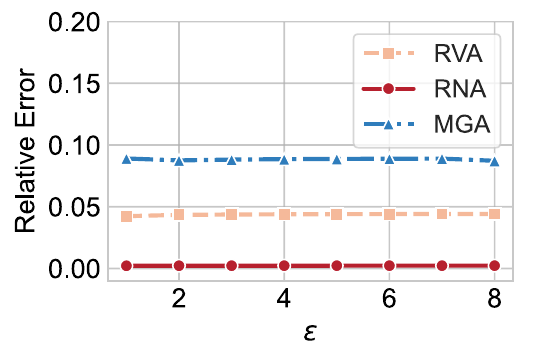}
        \caption{{\color{blue}LF-GDPR}}
    \end{subfigure}
    \hfill
    \begin{subfigure}[b]{0.49\linewidth}
        \centering
        \includegraphics[width=\linewidth]{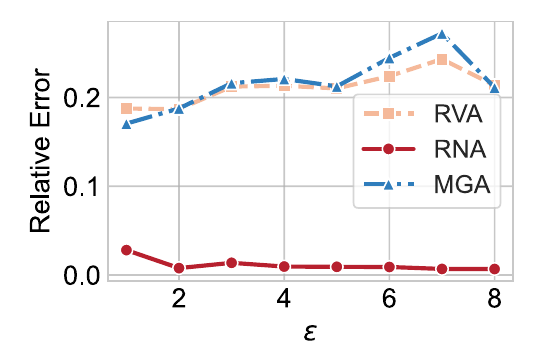}
        \caption{{\color{blue}LDPGen}}
    \end{subfigure}
    
    \caption{{\color{blue}Evaluation of attacks on LF-GDPR and LDPGen for modularity.}}
    \label{fig:evaluation_attacks_modularity}
\end{figure}

\subsection{{\color{blue}Results for Attacks on LF-GDPR and LDPGen}}
\noindent\underline{\textbf{{\color{blue}Exp 9. Evaluation of Attacks on LF-GDPR and LDPGen}}}. {\color{blue} We finally evaluate the performance of three attack strategies against LF-GDPR and LDPGen, focusing on their impact on clustering coefficient and modularity metrics \cite{ye2020,qin2017}. The results are shown in Fig. \ref{fig:evaluation_attacks_clustering_coefficient} and \ref{fig:evaluation_attacks_modularity}. For both metrics, all three attacks prove effective across various privacy budgets $\varepsilon$, with MGA generally achieving the best performance, followed by RVA and RNA. These findings confirm the efficacy of our proposed attack strategies.}

\section{Conclusion}\label{sec:conclusion}
In this work, we demonstrate the vulnerability of LDP protocols for graph data collection to data poisoning attacks. Despite the privacy guarantees of LDP, we show that attackers can inject fake users and launch targeted data poisoning attacks that significantly degrade the quality of collected graph metrics, such as degree centrality and clustering coefficients. We introduce three novel data poisoning attacks, and theoretically prove that our MGA attack can maximally distort graph metrics for targeted nodes. Our experimental evaluation on real-world datasets confirms the effectiveness of these attacks. We also explore two countermeasures but find them insufficient to mitigate the attacks' impacts. This underscores the urgent need for more robust defenses to secure LDP protocols for graph data collection and analysis.

\section{Acknowledgment}
This work was supported by the National Natural Science Foundation of China (Grant No: 62102334, 92270123, and 62072390), the Research Grants Council, Hong Kong SAR, China (Grant No:  15226221 and 15209922), and General Research Grants (Grant No: FRG-24-027-FIE) of the MUST Faculty Research Grants (FRG).

\vspace{12pt}

\end{document}